\documentstyle[12pt,a4]{article}
\begin{document}

\title{On Functional Determinants of Laplacians \\
in Polygons and Simplices}
\author{Erik Aurell $^1$, Per Salomonson $^2$\\\\
$^1$ Department of Mathematics,
University of Stockholm \\
S--113 85 Stockholm,
Sweden \\
$^2$ Institute of Theoretical Physics, Chalmers University of Technology\\
S--412 96 G\"oteborg, Sweden}
\maketitle

\begin{abstract}
The functional determinant of an elliptic operator with positive, discrete
spectrum may be defined as $e^{-Z'(0)}$, where $Z(s)$, the zeta function,
is the sum $\sum_n^{\infty} \lambda_n^{-s}$ analytically continued to
$s$ around the origin. In this paper $Z'(0)$ is calculated for the Laplace
operator with Dirichlet boundary conditions inside polygons and
simplices with the topology of a disc in the
Euclidean plane.
The domains we consider are hence piece--wise flat with corners on the
boundary and in the interior.
Our results are complementary to
earlier investigations of the determinants on
smooth surfaces with smooth boundaries.
The computation takes the form of a variation of the shape of the
domains, which is chosen such that the coordinate transformations
are conformal everywhere but at the corners. The contributions to
the variation of $Z'(0)$ in simplices
then come almost exclusively from terms with singular
support at the corners: in polygons there are no contributions
but from the corners.
We have explicit closed integrated
expressions for triangles
and regular polygons.
Among these, there are five special cases (three triangles,
the square and the circular disc), where the $Z'(0)$ are known
by other means. One special case fixes an integration constant,
and the others provide four independent analytical checks.
\end{abstract}

\bigskip
\centerline{\it Preprint, April 1993}
\newpage

\section{Introduction}
One of the basic integrals that arises in many parts in physics is
\begin{equation}
\label{gaussint}
\int_{-\infty}^{\infty}\prod_{k=1}^n {{dx_k}\over{(2\pi)^{1\over 2}}}
 e^{-{1\over 2}xA  x}
= (Det A)^{-{1\over 2}}
\end{equation}
where $A$ is a real, symmetric matrix with posive eigenvalues.
For instance, let (\ref{gaussint}) describe the integration
of fluctuations around a classical solution in quantum mechanics, where
the Lagrangian has been expanded up to second order.
The dimension of $A$ is then infinite, and (\ref{gaussint}) is
divergent on
both sides. The equation is therefore undefined as it stands.
As a basic example take
a one--dimensional
harmonic potential, and the fluctuations
in the time interval $0<\tau<L$. Then
\begin{equation}
xAx = \int_{0}^{L} d\tau (\partial_{\tau}x)^2 + \omega^2 x^2,
\end{equation}
and we have Dirichlet
boundary conditions for $x$
at $\tau=0$ and $\tau=L$.
The most straightforward way, is then to
go back to the Gaussian integral (\ref{gaussint}),
reintroduce a cut-off $\epsilon$
in space,
and modify the integration
measure depending on the cut-off so that the limit when $\epsilon$
goes to zero is finite.
In quantum mechanics this is feasible:
changing $(2\pi)^{-{1\over 2}}$ to $(2\pi\epsilon)^{-{1\over 2}}$,
and including one more factor $(2\pi\epsilon)^{-{1\over 2}}$,
turns (\ref{gaussint}) to  a discrete approximation to
Feynman's sum over paths, which in the limit
gives
\begin{equation}
\label{sinh}
(\hbox{Regularized[$Det A$]})^{-{1\over 2}} = ({{2\pi\sinh (\omega L)}\over
L})^{-{1\over 2}},
\end{equation}
and this is the correct expression in the Greens function.

We may also observe that
the eigenvalues of $A$ are
${{\pi^2 n^2}\over{L^2}} + \omega^2$, and
the determinant is then formally
\begin{eqnarray}
\label{divergdet}
\det A \sim \prod_{n=1}^{\infty} ({{\pi^2 n^2}\over{ L^2}} + \omega^2)
\end{eqnarray}
One way to regularize the determinant is to introduce a
cut-off $\Lambda$ in the product (\ref{divergdet}),
check that in the limit of large $\Lambda$ the result separates
into one finite factor and one factor divergent with
$\Lambda$, and keep the finite factor
as the regularized result.
For (\ref{divergdet}), this gives the same result as
(\ref{sinh})\cite{Parisi}.

A regularization can also be found from
zeta function of the operator;
\begin{equation}
\label{zetadef}
Z_A(s) = \sum \lambda_k^{-s}
\end{equation}
which converges when the real part of $s$ is large enough. When
this function can be analytically continued to
be regular in a neighbourhood
of the origin, then
\begin{equation}
\label{zetaprimedef}
\hbox{Regularized[$Det A$]} = e^{-Z_A'(0)}
\end{equation}
For (\ref{divergdet}), this again gives the same result as
(\ref{sinh}).
The zeta function method
was first introduced in the context of regularizing
expressions like (\ref{gaussint}) by Hawking\cite{Hawking},
to study fluctuating fields in a
background of curved space.

In dimension higher than one, it is not evident that all
regularizations of the determinant give the same result.
In this paper we will compute $Z_A'(0)$, with $A$ the
Laplace operator with Dirichlet boundary conditions in two--dimensional
simplicial domains.
In the language of field theory this is
the gaussian model, a free massless theory, albeit with unusual and
obstructing boundary conditions. Even so, the direct limit
is far from trivial, and the most relevant result we are aware of,
works only for lattice Laplacians, discretized
on ectangular ($M\times N$) domains\cite{DD}:
\begin{eqnarray}
\label{DDasympt}
\det A \sim 2^{9\over 4} e^{ {GMN}\over{\pi} }
(1+\sqrt 2 )^{ -{{(M+N)}\over 2} } (MN)^{-{1\over 4}}\eta (q)
({M\over N})^{1\over 4}
\end{eqnarray}
Here, in lattice units, $MN$ is the area, $2(M+N)$ is the length
of the boundary, $q=e^{-2\pi {N\over M}}$ is the modular parameter,
G is Catalan's constant and $\eta(q)$ is the modular form of
Dedekind. There are now no less than three terms separately
diverging with the size of the lattice.
If, with hindsight, we use that
for rectangular domains
$Z_A(0) = {1\over 4}$, we can rewrite (\ref{DDasympt}) in terms of
an explicit lattice spacing $a$:
\begin{eqnarray}
\label{DDsimple}
\det A \sim \mu_{A}^{{Area}\over{a^2}}\mu_{L}^{{Length}\over{a}}
(a^2)^{Z_A(0)} e^{-Z_A(0)\log Area - B}
\end{eqnarray}
where $e^{-B}$ are the various remaining terms in
(\ref{DDasympt}) which
agree\footnote{up
to a constant factor $2^{{11}\over{4}}$} with
$e^{-Z_A'(0)}$\cite{ItzZub2torus}({\it see} Appendix 11).
If nothing else, it seems likely that a discretization on a
rectangular grid, of a domain which is not itself of rectangular
shape, will give rise to oscillating terms in the cut--off.
If (\ref{DDsimple}) is to be generally valid in two
dimensions, it can probably only be of a smoothened discretized
determinant, where
the smoothening goes over cut--off scales.
Assuming that this can be done, and
considering that the area, the length of the
boundary, and $Z_A(0)$ are all integrals of local distributions
({\it see} section 2), it is possible to introduce local
cut--off dependent counter--terms, such that the
the finite remaining piece is $e^{-Z_A'(0)}$.

It therefore at least makes sense to define the regularized
determinant to be $e^{-Z_A'(0)}$, and this is the view we
take in the rest of this paper. We will use the notation
$Z_D'(0)$ for our generic case: the zeta function of the Laplacian with
Dirichlet boundary conditions in a simplicial domain $D$,
with the topology of a disc. We will freely change the index of
the zeta function to denote various special cases, and even
contributions to the regularized determinant from some parts
of the domain.

If we look for physical relevance, we must have fluctuating
geometry, as otherwise all would dissappear in an overall normalization.
It is a quite old idea that the elementary excitations of
non--Abelian gauge theories are string--like
objects\cite{Wilson,Polyakov1,Alvarez}.
In lattice gauge theories, the statistical weight of a Wilson
loop, when a quark and an anti--quark are taken apart for
some time, is the area of the smallest area delimited by
the loop. A phenomenological model
of the excitations was proposed\cite{Polyakov1}, where the
statistical weight of a surface is its area, and
the path integral goes over imbeddings in $d$--dimensional
external space (``$x^{\mu}$''), and over internal two--dimensional
geometry (``$g^{ab}$''):
\begin{equation}
\label{Polyakovint}
Z \sim \int D[g^{ab}] D[x^{\mu}] \,
e^{-{1\over 2}[\int \sqrt{g} g^{ab} \partial_{a} x^{\mu}
\partial_{b} x^{\mu}]}
\end{equation}
Our computations are relevant to a part of the investigations
of (\ref{Polyakovint}). With $g^{ab}$ fixed, the integration over
$x^{\mu}$ is just a quadratic integral
with Dirichlet boundary conditions like (\ref{gaussint}).
Our calculations hence
gives the finite piece of this determinant.
As is well known, reparametrization
invariance of the action in (\ref{Polyakovint}), gives rise
to Faddeev--Popov determinants, which for smooth surfaces turn
out to be determinants of Laplacians acting on vector fields,
with modified Dirichlet boundary conditions. We have not
investigated these determinants, and we are not quite sure they
are relevant when
we consider piece--wise flat
surfaces with sharp corners.
At least in a class of simplices with fixed number of corners,
such a surface is its own model, and we would not have any more
reparametrization invariance.
In this respect, an approach closer to the simplicial
discretization of (\ref{Polyakovint})
would seem to be more appropriate\cite{KKM}.

Let us now see why it may be interesting to investigate
the determinants on simplicial disc--like domains.
{\it Smooth} disc--like manifolds with {\it smooth boundary}
can always be mapped conformally onto one another. If we
denote the conformal factor $\sigma(x)$, the base metric and
curvature by $\hat g$ and $\hat R$, and the base geodetic
curvature of the boundary by $\hat k$, a celebrated
result\cite{Polyakov1,Alvarez} says that
\begin{eqnarray}
\label{Liouvilleaction}
Z'_{D'}(0) =
Z'_{D}(0) - {1\over{4\pi}}
\int_{\partial D} d\hat s \hat n\cdot\partial\sigma
+ {1\over{6\pi}}
\int_{\partial D} d\hat s \hat k\sigma \nonumber \\
+{1\over{12\pi}}
\int_{D} d^2z
\sqrt{\hat g} [\hat g^{ab} \partial_{a} \sigma
\partial_{b} \sigma  + \hat R\sigma]
\end{eqnarray}
The integration constant can be computed from the upper
half sphere\cite{Weisberger}. If we
disregard the boundary terms, the equations of motion
of (\ref{Liouvilleaction}) are Liouville's equation,
and we refer to the combined action in (\ref{Liouvilleaction})
as the Liouville action.

In ordinary Feynman path integrals (\ref{gaussint}) the
typical path is quite rough, i.e. a nowhere smooth random walk.
It seems likely that the typical surface
that enters in \ref{Polyakovint} is also quite rough.
On two--dimensional smooth surfaces, one
can open up a corner with angle $2\pi\alpha$
(or $\pi\alpha$ at the boundary) with a coordinate transformation
which is conformal and regular everywhere but at the corner,
where it instead has a logarithmic singularity.
The kinetic energy term
in (\ref{Liouvilleaction}) will then be logarithmically
divergent at the corner.
In other words, the action for smooth (and conformal) coordinate
changes is infinite for these transformations.
One possible procedure is then to
introduce a cut-off $r_0$  at a corner,
and simply remove the quantities diverging
when $r_0$ goes to zero\cite{HP}. This is obviously dangerous;
if one is not careful one easily gets a spurious finite
piece from a logarithmic divergence, and
$Z_D'(0)$ is a  well-defined mathematical object, which
has a value, the surface being smooth or not.
The correct interpretation is that a conformal transformation
does not change $Z_D(0)$, but a non--conformal one
does. In fact, for piece--wise flat surfaces, $Z_D(0)$ can
be written as a sum over a rational functions of the
opening angles of the corners ({\it see} section 2, Appendix 6).
A typical non--conformal transformation will change the
opening angles by a shear, so it will change $Z_D(0)$.
The infinity of (\ref{Liouvilleaction}) under non--conformal
transformations is therefore a mirror of actually $Z_D(0)$ changing.

Our computation give some more explicit results on determinants,
to which one does not have access from smooth models. This
can have some mathematical interest by itself. More speculatively,
it is possible that a definition of determinants by a precise
calculation of $Z'_D(0)$, may yield a better regularization
of (\ref{Polyakovint})
than does (\ref{Liouvilleaction}) and its Faddeev--Popov
ghosts.
Certainly, such a result would go far beyond what is actually done
here: we have not begun to address a computation of a sum
over surfaces as in
(\ref{Polyakovint}).

The conclusions of this paper can now be stated as the following
propositions:
\\\\
{\bf Proposition 1.} {\it Domains with disc--like topology may
be mapped to the upper half complex plane, the boundary being
mapped to the real axis, and the corners being mapped to
branch--points $\omega_j$. If $z$ is the coordinate in the domain,
the transformation satisfies
\begin{eqnarray}
{{d\omega}\over{dz}}
= \phi_D\prod_j (\omega-\omega_j)^{1-\alpha_j},
\end{eqnarray}
and the representation is determined by the angles
and the lengths of the sides,
up to a rational fractional transformation of the upper half
plane. Choosing one parametrization
the normalized area of the simplex is:
\begin{eqnarray}
\hbox{Area}_D = \int_{\Im\omega > 0} {{d\bar\omega\wedge d\omega}\over{2i}}
\prod_j |\omega-\omega_j|^{2\alpha_j-2},
\end{eqnarray}
and the variation
of $Z_D'(0)$ under a general shear and dilatation
can be written;
\begin{eqnarray}
\delta[Z_D'(0)|_{\hbox{Area} = A}]
= \delta[Z_D'(0)] + \delta[Z_D(0)\log A] - \delta[Z_D(0)\log\hbox{Area}_D]
\end{eqnarray}
where then value of the zeta function at the origin is;
\begin{eqnarray}
Z_D(0) = \sum_{\hbox{interior corners}}{1\over{12}}({1\over{\alpha_j}}
		-\alpha_j)
+\sum_{\hbox{boundary corners}}{1\over{24}}({1\over{\alpha_j}} -\alpha_j)
\end{eqnarray}
and the opening angles are written $2\pi\alpha_j$ in the
interior, and $\pi\alpha_j$ on the boundary.
}
\\\\
{\bf Proposition 2.} {\it
When the area is chosen $\hbox{Area}_D$, the
difference of $Z_D'(0)$ between two
simplices, that differ by an infinitessimal
transformation, which is regular and conformal everywhere except
at the corners, can be written as a sum over quantities
with point mass support at the corners, and a line density on
the boundary. The integral over of the boundary term is identically
zero
for polygons, but gives
\begin{equation}
-4\pi\sum_{\hbox{interior corners}}
\delta\alpha_i,
\end{equation}
if there are corners in the interior.
}
\\\\
{\bf Proposition 3.} {\it The contribution to $\delta Z_D'(0)$ from
the point mass in one
corner $c$, is a sum
\begin{equation}
\delta Z_D'(0)|_{c} = \delta Z'_{\alpha_c}(0) +
Z_{\alpha_c}(0)\cdot\delta[\hbox{other corners}]
\end{equation}
where $\delta Z'_{\alpha_c}(0)$ is exclusively determined locally
at the corner, $Z_{\alpha_c}(0)$
is the contribution to $Z_{D}(0)$ from the corner,
and the influence from the
other corners depend
on the opening angles of these, and on the lengths of the sides:
\begin{equation}
Z_{\alpha_c}(0)\cdot\delta[\hbox{other corners}] = Z_{\alpha_c}(0)
[\sum_{c'\neq c}
\delta\alpha_{c'}\log |\omega_{c'}-\omega_c|^{2}
-(1-\alpha_{c'})
[{{\delta(\omega_{c'}-\omega_c)}\over{\omega_{c'}-\omega_c}} + \hbox{c.c}]]
\end{equation}
}
\\\\
{\bf Proposition 4.} {\it The strictly local contributions
may be given in integrated form, and are for a corner
on the boundary:
\begin{eqnarray}
Z'_{\alpha}(0) = {1\over{12}}({1\over{\alpha}} -\alpha)(\gamma -\log 2)
-{1\over{12}}({1\over{\alpha}} + 3 +\alpha)\log\alpha +\tilde J(\alpha) ,
\end{eqnarray}
and for a corner in the interior;
\begin{eqnarray}
Z'_{\alpha}(0) = {1\over{6}}({1\over{\alpha}} -\alpha)(\gamma -\log 2)
-{1\over{6}}({1\over{\alpha}} + \alpha)\log\alpha +2\tilde J(\alpha) ,
\end{eqnarray}
where the term $\tilde J$ has the integral representation;
\begin{eqnarray}
\label{Jdefmain}
\tilde J(\alpha) =
\int_{0}^{\infty}
{1\over{e^{x} -1}}
[{1\over {2x}}(\coth({x\over{2\alpha}}) -
\alpha \coth({x\over{2}}))
-{1\over {12}}({1\over{\alpha}}-\alpha)] dx.
\end{eqnarray}
We have here included an integration constant in our definition
of $\tilde J$.
}
\\\\
Special situations are quite important to us, as they provide
necessary checks. We therefore list the them separately:
\\\\
{\bf Proposition 5.} {\it
For triangles one may choose a parametrization where the
branchpoints lie fixed in $0$, $1$ and $\infty$. The normal
area of a triangle with
angles $\pi\alpha_1,\pi\alpha_2,\pi\alpha_{3}$ is then
\begin{equation}
Area(\alpha_1,\alpha_2,\alpha_{3}) = {{\pi}\over 2}
{{\Gamma(\alpha_1)\Gamma(\alpha_2)\Gamma(\alpha_{3})}\over
{\Gamma(1-\alpha_1)\Gamma(1-\alpha_2)\Gamma(1-\alpha_{3})}}
\end{equation}
and all terms but the strictly local in the variation of $Z'_D(0)$
vanish.
The determinant of a triangular domain is thus
\begin{equation}
Z_T'(0) = \sum_{p=1,2,3} Z'_{\alpha_p}(0)
\end{equation}
with $Z'_{\alpha}(0)$ as in Proposition 4.}
\\\\
{\bf Proposition 6.} {\it For regular polygons with $n$ corners
one can choose a parametrization by mapping to the unit disc,
where the corners are regularly spaced on the unit circle.
The radius of the circumscribed circle is with this parametrization:
\begin{equation}
R_n = {{\Gamma(1+{1\over n})\Gamma(1-{2\over n})}
        \over{\Gamma(1-{1\over n})}}
\end{equation}
and the determinant in a regular polygon with radius
of circumscribed circle $R$ is;
\begin{eqnarray}
Z_{P_n}'(0) = Z_{P_n}(0)\log{{R^2}\over{R_n^2}} -
{1\over{3(n-2)}}\log n + n Z_{1 - {2\over n}}'(0)
\end{eqnarray}
where $Z_{P_n}(0) = {{n-1}\over{6(n-2)}}$.
Specializing to $n=2$ we obtain the determinant in a square:
\begin{eqnarray}
Z_{P_4}'(0) = {1\over 4}\log\hbox{Area}
+{1\over 2}\log{{\Gamma({3\over 4})}\over{\Gamma({1\over 4})}}
+ {1\over 4}\log\pi + {5\over 4}\log 2
\end{eqnarray}
Taking the limit as $n\to\infty$ we obtain the determinant
in a disc:
\begin{eqnarray}
Z_{P_{\infty}}'(0) =
{1\over 3}\log R + {5\over 12} + {1\over 2}\log\pi + {1\over 6}\log 2
+ 2\zeta'(-1)
\end{eqnarray}
}
\\\\

The paper is organized as follows: standard results on
heat kernels and zeta functions
are summarized in section 2. In section 3 a representation is found
for $\delta Z'(0)$ which treats the corners and the rest in different
ways; this is the main idea. In both cases the representation
is in terms of the short--times asymptotics of the heat kernel.
In the interior this representation is standard, and gives
the variational form of (\ref{Liouvilleaction}). That the
underlying space is flat, and with disc--like topology,
actually simplifies things so much that this contribution
gives only the simple sum in Proposition 1.
In the corners the appropriate asymptotics is the
Sommerfeldt
heat kernel in an infinite sector.
Most of the derivations are by calculation, and most
are elementary (although sometimes
cumbersome)
We have chosen to present those in appendices, largely in a self--contained
way, but without any reference to physical
arguments. The appendices can therefore be read
more or less independently
from the rest of the paper.
Section 4 summarizes
the results.

\section{Heat kernel and Zeta function phenomenology}
The diagonal elements of the heat kernel on smooth manifolds
admit an asymptotic expansion for short
times. In two dimensions the expansion goes as
\begin{equation}
\label{asympt}
K_D(x,x,t) \sim {{c_{-1}(x)}\over t} +
{{c_{-{1\over 2}}(x)}\over {t^{1\over 2}}} + c_0(x) + \ldots
\end{equation}
McKean and Singer\cite{McKean} proved the existence of this
expansion for smooth manifolds up to terms $t$, and for manifolds
with smooth boundary up to $t^{1\over 2}$. In the interior
the coefficients $c_i$
are polynomials in the curvature tensor,
and on the boundary they are line distributions, with weight
depending on the curvature of the boundary.
For polygonal domains in the plane, Kac\cite{Kac} proved the expansion up
to $c_0(x)$, which in this case is made up of point masses at the corners.
Since
the curvature of a polygon can be said to be concentrated at the corners,
the result is in some sense natural,
and $c_0(x)$ for a smooth boundary can indeed
be found as the limiting case of
polygons with angles coming closer and closer to $\pi$.
However, the converse is
not true, that is, if one wants to compute $Z_D(0)$ for a manifold with
tips and corners, it is not possible to use a smoothened
approximation.

The integrated form of (\ref{asympt}) is
\begin{equation}
\label{asymptint}
\int_D dx^2 K_D(x,x,t) = \Theta_D (t) \sim {{c_{-1}}\over t} +
{{c_{-{1\over 2}}}\over {t^{1\over 2}}} + c_0 + \ldots
\end{equation}
The zeta function is defined by a Mellin transform of the trace of the
heat kernel as
\begin{eqnarray}
Z_D(s) = {1\over{\Gamma(s)}}
\int_0^{\infty} dt t^{s-1} \Theta_D (t)
\end{eqnarray}
The integral only converges in the lower limit if $\Re s>-\alpha_i$ for
all terms $c_{\alpha_i}t^{\alpha_i}$  in the heat kernel, i.e.
$\Re s>1$ in two dimensions.
One can get around the pole by a partial integration, where the
boundary terms vanish for $\Re s>1$:
\begin{eqnarray}
{1\over{\Gamma(s)}}
\int_0^{\infty} dt t^{s-1} \Theta_D (t) =\nonumber \\
{1\over{\Gamma(s)}}
[({{t^{s-1}}\over{s-1}}  t \Theta_D (t))]_0^{\infty}
- {1\over{\Gamma(s)}}
\int_0^{\infty} dt {{t^{s-1}}\over{s-1}}{{\partial}\over{\partial t}}
[t \Theta_D (t)]
\end{eqnarray}
The second integral now gives an analytic continuation down to
$\Re s>{1\over 2}$.
Further on the zeta function has poles at
$s=-\alpha_i$, with residue
${{c_{\alpha_i}}\over{\Gamma(\alpha_i)}}$, except at zero and
the negative integers, where the inverse gamma function has zeros,
which gives finite values $Z(-n) = (-)^nc_{-n}n!$\cite{BSV}.

For positive  $n$ greater than 1, a method that goes back to
Euler\cite{Watson}
gives a representation
of the zeta function as a convolution of electrostatic
Greens functions:
\begin{eqnarray}
Z_D(n+1) = Tr_x ({1\over{-\Delta}})^{n+1} = \int dx\int dy_1 \ldots \int dy_n
G_D(x,y_1)G_D(y_1,y_2)\ldots G_D(y_n,x)
\end{eqnarray}
This method can be extended to calculate the finite part of $Z_D(s)$ at
$s=1$\cite{IML}. Hence, if the  expansion (\ref{asymptint}) goes
on indefinitely and one knows the Greens function,
one may in principle calculate the
zeta function at the integers, and the residues at all the poles.
But if this is true, and the zeta function does not grow
too fast at infinity, then the zeta function is completely
determined, which (somewhat indirectly) determines the
eigenvalues. Hence all the information about the
eigenvalues, and on all quantities depending on them,
is then contained in the Green's function and the
asymptotic expansion of the heat kernel for short times.

It is natural to consider what we will call
a {\it zeta function density}:
\begin{eqnarray}
\label{zetalocal}
Z_D(x,x,s) = {1\over{\Gamma(s)}}
\int_0^{\infty} dt t^{s-1} K_D(x,x,t)
\end{eqnarray}
For $\Re s<1$  we define the zeta function density by analytic
continuation from (\ref{zetalocal}).
At zero, the zeta function density is determined by the
asymptotics of the heat kernel for short times as
in (\ref{asympt}):
\begin{eqnarray}
\label{zeta0finite}
Z_D(x,x,0) = c_0(x)
\end{eqnarray}
We may just as well consider the density associated
to the derivative of the zeta function at the origin,
but this quantity will depend also on
the heat kernel for long times. These questions are addressed
in more detail in Appendix 1.

One could also look at
off--diagonal elements in (\ref{zetalocal}), or
equivalently, the operator $({1\over{-\Delta}})^s$.
Since the off--diagonal Mellin transforms are finite
for all $s$, the analytical continuation brings
no off--diagonal subtractions, and effectively the diagonal and
off--diagonal elements are treated differently for $\Re s<1$.
It would be interesting to have a proper regularization
of off--diagonal terms, so that one might consider
$Z_D'(0)$ as the trace of a (regularized)
operator $Q\sim -\log(-\Delta)$,
and in terms of which a regularized Laplace operator
would be $e^{-Q}$ (in the operator sense).
In this paper we will only consider the diagonal
elements, and we therefore refrain from calling
the zeta function density the matrix elements of an operator
when $\Re s < 1$.

Let us consider a dilation
$x\to\lambda x$ of the domain. That changes the eigenvalues
in a simple way: $E_n\to\lambda^{-2}E_n$, and the zeta function
accordingly also changes simply: $Z_{\lambda D}(s)\to\lambda^{2s}Z_{D}(s)$.
For the quantities arond the origin this implies:
\begin{eqnarray}
\label{zetadil}
Z_{\lambda D}(0)=Z_{D}(0)\qquad Z_{\lambda D}'(0)=Z_{D'}(0)+\log\lambda^2
Z_{D}(0)
\end{eqnarray}
The invariance of $Z_{D}(0)$ under dilations is a special case
of invariance under regular conformal transformations. The
change in $Z_{D}'(0)$ logarithmically proportional to the variation
of the area, is contained in Proposition~1,
and was already noticeable in the asymptotic expansion
of the discretized determinant in rectangular domains
(\ref{DDsimple}).

\section{A variational formula}
In this section we will write down a variational formula.
It can be formulated either in terms of the asymptotics of
the heat kernel for short times, or in terms of the zeta function
density from section 2.
Let us first motivate why we give two computations, that essentially
only differ in that they work on opposite sides of the
Mellin transform: the approach using the heat kernel is more
traditional, but requires more delicate treatment to separate
out the finite piece. Note that our formulae also differ by
(a fairly simple term) from a definition of the regularized
determinant often used in field theory\cite{Alvarez,Polyakov1}.
The reason for this discrepancy is that we also consider
transformations that are not conformal, and here the
additional term matters. The approach using the zeta function is
computationally somewhat simpler, once the notion of a zeta
function density is admitted.

Let now $D$ and $D'=D+\delta D$ be two domains that differ infinitesimally.
Then, for $\Re s>1$,
\begin{equation}
\label{varyT}
Z_{D'}(s)-Z_D(s) = \sum_{n=0}^{\infty}[E_n'^{-s} - E_n^{-s}] =
-s\sum_{n=0}^{\infty}({{\delta E_n}\over{E_n}})E_n^{-s}
\end{equation}
and this definition is analytically continued to lower values of
$s$.
We can consider $\{E_n'\}$ to be the eigenvalues of a
modified operator
$(-\Delta-\delta\Delta)$ in the domain $D$, so that we write
\begin{equation}
\label{smallvaryT}
Z_{D'}(s)-Z_D(s) = s Tr_x [(\delta\Delta)  Z_D(x,x,s+1)]
\end{equation}
The variations we will consider map one simplex on another. They
therefore leave the following sum invariant:
\begin{equation}
\label{invariant}
{\cal S}(\alpha_j) = + \sum_{\hbox{boundary corners}} (1-\alpha_j) +
\sum_{\hbox{interior corners}} (2-2\alpha_j)
 =2
\end{equation}
where the opening angle of a corner on the boundary
(in the interior) is $\pi\alpha$ ($2\pi\alpha$).
The analytical expression for our variation
will therefore only be determined up to the variation of
a ($s$--dependent)
function of ${\cal S}(\alpha_i)$,
that is, up to terms linear in the variations of the
angles.

If we want to evaluate the variation of the derivative
at zero we can do it as
\begin{equation}
{d\over {ds}}\delta Z_{D}(s)|_{s=0} =
\lim_{s\to 0}
{1\over s}Tr_x [ (Z_{D'}(x,x,s) - Z_{D}(x,x,s))
- (Z_{D'}(x,x,0) - Z_{D}(x,x,0))]
\end{equation}
which can be rewritten as
\begin{equation}
\label{finitevary}
{d\over {ds}}\delta Z_{D}(s)|_{s=0} =
\hbox{Finite}_{s\to 0}
{1\over s}Tr_x [ Z_{D'}(x,x,s) - Z_{D}(x,x,s)]
\end{equation}
In general there will be a pole at the origin in $s$
in (\ref{finitevary}), with a residue equal to the
variation of $Z_D(0)$. Combining (\ref{finitevary})
and (\ref{smallvaryT}) we have
\begin{equation}
\label{smallfinitevary}
{d\over {ds}}\delta Z_{D}(s)|_{s=0} =
\hbox{Finite}_{s\to 0}
Tr_x [(\delta\Delta) Z_D(x,x,s+1)]
\end{equation}
It is now convenient to choose the variation
in a particular way.
Think first
of the Laplace--Beltrami operator in
curved space, with metric tensor $g^{ab}$:
$\Delta = g^{-{1\over 2}}
\partial_a g^{ab}g^{{1\over 2}}\partial_b$.
In two dimensions we can choose the particular form
$g_{ab} = e^{2\sigma(x)} \hat g_{ab}$, and for a
smooth surface with disc
topology we can take $\hat g^{ab}$ to be the standard
flat metric. In this coordinate system, the
Laplacian has the following form
$\Delta = e^{-2\sigma}\partial^2_{aa}$.
A conformal variation of the Laplace--Bertrami operator
is then $\delta\Delta = (-2\delta\sigma)\Delta$, and
the final variational formula is
\begin{equation}
\label{varyPERS}
{d\over {ds}}\delta Z_{D}(s)|_{s=0} =
\hbox{Finite}_{s\to 0}
Tr_x [2\delta\sigma(x) Z_D(x,x,s)]
\end{equation}
The formula (\ref{varyPERS}) holds also
when we include
corners with varying opening angles,
the only new effect then being that prefactor
weight $\delta\sigma$ has a logarithmic
singularity at the corner.

We will show in Appendix 3 that in flat domains far
from the boundary, the density $Z_D(x,x,s)$ vanishes
at the origin. Hence the interior of flat domains,
far from the corners, give no contribution at all
to $Z_D'(0)$. This is in agreement with
(\ref{Polyakovint}), since our base space is flat (hence
$\hat R$ is zero), and the kinetic energy term of a regular
conformal variation can only give a boundary term.
At straight boundaries, there is
an undetermined
contribution to $Z_D'(0)$, proportional to the boundary length.
The contribution to the variation, $\delta Z_D'(0)$, is however
fully determined following (\ref{varyPERS}), and picks out the normal
derivative of the variation $\delta\sigma(x)$. Here
again we have exact agreement with (\ref{Liouvilleaction}).

For simplices, the boundary term integrates trivially,
and we are left with only the corner contributions as the
important parts.
Considering that the zeta function density
$Z_D(x,x,0)$, has components with point mass support
at the corners, it may be surmised that integrated
against the logarithm of the distance from the tip,
it gives an infinite contribution to
(\ref{varyPERS}). This is indeed the case, that
infinity is precisely the variation of $Z_D(0)$,
which we subtract by taking the finite part as
$s$ goes to zero. It is now important to realise
that we can choose small areas around the corners,
with some radius $r_0$, which we can let tend to zero
at the end of the calculation. It therefore does not matter
if we compute the answer up to terms proportional to
$r_0$ or $r_0^2$, since these will eventually drop out.
This means that we can actually substitute the true (unknown)
$Z_D(x,x,s)$, with a sufficiently good approximation,
computed from the Mellin transform of an approximation
to the true heat kernel valid for short times.
The technique for doing this is explained in Appendix~5,
and the approximation to the heat kernel -- the
Sommerfeldt heat kernel in a sector -- is described
in Appendix~4.
The calculations of the contributions
to $Z_D'(0)$ from the corner is then
done in Appendix~6.

Let us now derive an alternative formula to
(\ref{varyPERS}) using only the heat kernel for
short times. We begin with the following
equality (derived in Appendix 1):
\begin{equation}
\label{AlvarezcorrR}
Z_D'(0) =
\gamma Tr_x Z_D(x,x,0)
+ \hbox{Finite}_{\epsilon\to 0} Tr_x
\int_{\epsilon}^{\infty} {{dt}\over t}  K_D(x,x,t)
\end{equation}
The heat kernel can be expanded in a complete set of states:
\begin{equation}
K_D(x,x,t) = \sum_v |\psi_v(x)|^2 e^{-\lambda_v t}
\end{equation}
and the
variation of (\ref{AlvarezcorrR}) will be
\begin{eqnarray}
\label{AlvarezcorrR2}
\delta Z_D'(0) =
\gamma \delta Tr_x Z_D(x,x,0)
-\hbox{Finite}_{\epsilon\to 0}
\int_{\epsilon}^{\infty} dt
\sum_v <v|(2\delta\sigma\Delta) |v> e^{-\lambda_v t}  \nonumber \\
= \gamma \delta Tr_x Z_D(x,x,0)
+\hbox{Finite}_{\epsilon\to 0}
 Tr_x [2\delta\sigma(x) K_D(x,x,\epsilon)]
\end{eqnarray}
When we are in the interior, or at a straight boundary,
the zeta function density $Z_D(x,x,0)$ is zero.
The remaining trace over the heat kernel in (\ref{AlvarezcorrR2})
then agrees with a well--known definition of the regularized
determinant in field theory, and
the short--time expansion of the heat kernel can be done
by standard means\cite{Polyakov1,Alvarez}.
We then find the
variational form of (\ref{Liouvilleaction}).
In piece--wise flat simplices there are further simplifications,
such that the only non--zero term is the integral over
the boundary of the normal derivative of $\delta\sigma$ in
(\ref{Liouvilleaction}).

Close to the corners, we cannot
however use the same short--time expansion of the heat kernel.
Again we have to turn to the Sommerfeldt heat kernel,
and carefully extract the constant piece of the trace
of the heat kernel
for short times. Finally, we must
take into consideration that the variation of $Z_D(0)$ is not zero
at the corner, and add the first term on the right--hand side of
(\ref{AlvarezcorrR2}). The computation following these lines
is done in Appendix~7.

\section{Calculations \& Conclusions}
In this section we describe the computations
in the appendices, and discuss the results.

Appendix~1 is an expansion on Section~2. We establish that
the zeta function density of the derivative at zero is well--defined,
but depends on the heat kernel for long times. An
alternative definition of the variation of the regularized
determinant in terms on the heat kernel for short times
is derived. Appendix~2 presents a general
parametrization of variation of simplices, and
computes the contribution from corners in the
interior as in Proposition~2. In Appendix~3 we treat in some
detail the case of a straight boundary. We derive that it
cannot give a a term in $Z_D'(x,x,0)$ more singular than
a line density, and hence a contribution to $Z_D'(0)$
proportional to boundary length. We also derive the
contribution from the line integral on the boundary to
the variation of $Z_D'(0)$. Appendix~4 presents
the Sommerfeldt heat kernel in a sector, and Appendix~5
states an elementary integral, that is convenient when
one wants to compute the Mellin transform using the Sommerfeldt
heat kernel.

In Appendix~6 we compute the contribution to $Z_D(0)$, and
the strictly local contribution to
$Z_D'(0)$ from a corner with opening angle $\alpha$,
using the approach of the zeta function density.
We call these quantities $Z_{\alpha}(0)$ and $Z_{\alpha}'(0)$.
We perform the computations for both corners on the boundary
and corners in the interior. In Appendix~7 we do the same
thing, using the slightly more involved approach of the
analysis of the short--time properties of the Sommerfeldt heat
kernel. In Appendix~8 we investigate $Z_{\alpha}'(0)$ in
the special cases,
where the opening angle is ${1\over n}$. In Appendix~9
we investigate the asymptotic behaviour of $Z_{\alpha}'(0)$
as $\alpha$ is large, small or close to one.

In Appendix~10 we treat a few special cases using our general
variational formula. The additional computations done here
essentially boil
down to taking care of the expression in Proposition~3,
that has support at one corner, but depends on the
angles of all the corners, and on the lengths of the
sides. In the very special case of triangles, this
additional term is absent, but in polygons it is important.
In Appendix~11 we state summarily the cases we know to
compare with,
where the $Z_D'(0)$ are known, either because the eigenvalues
are known, or because these domains are conformal images
of some domain where the eigenvalues are known. This
is important information: it fixes an integration
constant that we cannot reach; and it provides
much needed analytical checks.

In conclusion we have shown that the determinants on piece--wise
smooth surfaces are different from the determinants on
smooth surfaces. The formula that describe variations
of determinants on smooth surfaces (\ref{Liouvilleaction})
give infinite results for variations that change the
opening angles of the corners. This infinity is not real,
it is only an imprint of actually $Z_D(0)$ changing.
We have parametrized the variations of the domains
such they are
conformal everywhere outside the corners,
and separated the contributions to the variation
of $Z_D'(0)$ into four parts:
one (Proposition~1), that expresses the variation of
$Z_D(0)$ times the logarithm of the area;
one (Proposition~2), which
is an integral over the
boundary of the normal derivative of the
conformal parameter; one (Proposition~3)
which has local support at the corners, but depend
on the opening angles of all the corners and on the lengths
of the sides; and one (Proposition~4),
which is determined strictly locally at the corners.

The first part amounts to determining how the normalized
area changes with the angles and the sides.
We can trivially integrate the second part, which is
identical to one
of the terms in the variation of determinants on smooth surfaces
(\ref{Liouvilleaction}).
We have integrated the fourth part, both for corners
on the boundary and in the interior.
For general opening angles this part has an integral
representation (\ref{tibc}, \ref{Jdef2}), which for the
special opening angles $\alpha={1\over n}$, can
be resolved into a finite sum (\ref{finalfixed}).

We have been able to parametrize the area and integrate
the third part, for the special cases of triangles
(where the third part vanishes) and a class of polygons,
which includes certain interpolations between pairs of
regular polygons (Appendix~10).
In general, we have not been able to integrate the third part,
which therefore has to be left in variational
form (Proposition~3).

The wider applicability of our approach evidently
depends on better handling of the parts we have
not integrated.
We do not expect that one will in general
be able  even to express the normalized area in closed
form, as a function of the angles and the sides.
Perhaps one may however find a limiting form, which
is valid for small opening angles.
Considered as an action for simplicial surfaces,
it is possible that the full expression of $Z_D'(0)$
strongly damps out corners with small opening
angles.
As is well known, most investigations of sums
over random surfaces in physical dimensions
lead to very rough surfaces, dominated by long
sharp spikes.
If the full expression of $Z_D'(0)$ as a function
of one opening angle has a singularity at the
origin -- with the right sign of the prefactor --
then surfaces with spikes will be more strongly
damped by the action $Z_D'(0)$, then by any finite
expansion in local curvature.
We hope to return to questions in this direction
in the future.
\section{Acknowledgements}
E.A. wishes to thank
Claude~Itzykson for initially encouraging
him to study these problems, and
Service de Physique Th\'eorique at Saclay
for hospitality in 1986--87. We thank Jean--Marc Luck for communicating
numerical results on the regularized determinants in triangles,
that proved a valuable check.
This work was supported by the
Swedish Natural Science Research Council under contracts
U--FR--1778--101, S--FO~1778--302 and F-FU 8230-306,
and by the G\"oran Gustavsson Foundation.
\newpage

\section* {A1. Local and nonlocal quantities}
\label{s:app1}
\renewcommand{\theequation}{A1.\arabic{equation}}
\setcounter{equation}{0}
In this appendix we collect some results on the zeta
function density and compare it to other approaches in
the literature. An important aspect is locality: a Mellin
transform of a heat kernel, at a point, integrated over some
subdomain, or over the whole domain, is said to be local,
if it is determined only by the expansion of the heat kernel
for short times.

We start with the definition (\ref{zetalocal})
\begin{equation}
Z_D(x,x,s) = {1\over{\Gamma(s)}}
\int_0^{\infty} dt t^{s-1} K_D(x,x,t)
\end{equation}
It is useful to explicitly analytically continue the integral beyond
the poles at $s=1$, $s={1\over 2}$ and also beyond $s=0$. We will
then get out a pole in $s$ at $s=0$ from the integral,
which will combine with the gamma--function
in front to give a regular function. We thus have a representation of
the zeta function density around $s=0$ in terms of a regular prefactor
and a convergent integral:
\begin{equation}
Z_D(x,x,s) = {1\over{(s-1)(s-{1\over 2})s\Gamma(s)}}
\int_0^{\infty} dt t^{s}
{{\partial}\over{\partial t}}[
t^{1\over 2}{{\partial}\over{\partial t}}[t^{1\over 2}
{{\partial}\over{\partial t}} t K_D(x,x,t)]]
\end{equation}

Using the asymptotic expansions as $t$ is close to zero:
\begin{eqnarray}
K_D(x,x,t) \sim {{c_1(x) }\over t} +
{{c_{1\over 2}(x) }\over {t^{1\over 2}}} +
c_0(x) + c_{-{1\over 2}}(x) t^{1\over 2} + \ldots \nonumber \\
{{\partial}\over{\partial t}} t K_D(x,x,t) \sim
{1\over 2}{{c_{1\over 2}(x) }\over {t^{1\over 2}}} +
c_0(x) + {3\over 2} c_{-{1\over 2}}(x) t^{1\over 2} + \ldots \nonumber \\
t^{1\over 2}[{{\partial}\over{\partial t}}[ t^{1\over
2}{{\partial}\over{\partial t}} t K_D(x,x,t)] \sim
{1\over 2}c_0(x) +
{3\over 2} c_{-{1\over 2}}(x) t^{1\over 2}+ \ldots  \nonumber
\end{eqnarray}
we may evaluate as follows
\begin{eqnarray}
\hbox{Res} Z(x,x,s) |_{s=1} =&  c_1(x) \nonumber \\
\hbox{Res} Z(x,x,s) |_{s={1\over 2}} =&  \pi^{1\over 2}c_{1\over 2}(x)
\nonumber \\
Z(x,x,0) =& c_0(x) \nonumber
\end{eqnarray}

We may also evaluate the derivative with respect to $s$ at
the origin:
\begin{equation}
\label{zetaprimdef}
Z_D'(x,x,0) =
(3+\gamma)c_0(x) +
2\int_0^{\infty} dt \log t
{{\partial}\over{\partial t}}[
t^{1\over 2}{{\partial}\over{\partial t}}[t^{1\over 2}
{{\partial}\over{\partial t}} t K_D(x,x,t)]]
\end{equation}
It is quite clear that $Z_D'(x,x,0)$ is a quantity
which depends also on the heat kernel at large times.

An alternative formula is obtained by taking
the integral from
$\epsilon$ to infinity, where $\epsilon$ is some small positive
number.
As the integrand in \ref{zetaprimdef}
behaves as $t^{-{1\over 2}}\log t $
for small $t$, the integral is convergent, and the limit
as $\epsilon$ goes to zero is harmless. Partial integrations
will however bring in divergent terms from the lower
boundary. A short calculation gives
\begin{eqnarray}
Z_D'(x,x,0) =
(3+\gamma)c_0(x) + \nonumber \\
(c_0(x)\log\epsilon  - 2c_{1\over 2}(x) \epsilon^{-{1\over 2}}
-c_1(x)\epsilon^{-1} - 3 c_0(x))
+\int_{\epsilon}^{\infty} {{dt}\over t}  K_D(x,x,t)
\end{eqnarray}
which can be simplified to
\begin{equation}
\label{Alvarezcorr}
Z_D'(x,x,0) =
\gamma c_0(x)
+ \hbox{Finite}_{\epsilon\to 0}
\int_{\epsilon}^{\infty} {{dt}\over t}  K_D(x,x,t)
\end{equation}
Let $\tilde K_D$ be an asymptotic approximation to $K_D$, valid
for short times. We may then write the integral in
\ref{Alvarezcorr} as
\begin{eqnarray}
\label{intdivided}
\hbox{Finite}_{\epsilon\to 0}\int_{\epsilon}^{1} {{dt}\over t}  \tilde
K_D(x,x,t)
+ \int_{0}^{1} {{dt}\over t}  [K_D(x,x,t) - \tilde K_D(x,x,t)]
+ \int_{1}^{\infty} {{dt}\over t}  K_D(x,x,t)
\end{eqnarray}
The second and third terms are finite, if the asymptotic
approximation is sufficiently good. Hence they
can only contribute to $Z_D'(x,x,0)$ as a smooth density.
The only term which can give a distribution contributions,
the space integral of which contains a finite piece even if
taken over a vanishingly small area,
is the first in \ref{intdivided}, and the term $\gamma c_0(x)$. Both
only depend on the heat kernel for short times.

When we integrate \ref{Alvarezcorr} over the whole domain,
we obtain
\begin{equation}
Z_D'(0) =
\gamma Z_D(0)
+ \hbox{Finite}_{\epsilon\to 0} Tr_{x}
\int_{\epsilon}^{\infty} {{dt}\over t}  K_D(x,x,t)
\end{equation}

\section* {A2. Variation and parametrization}
\label{s:app2}
\renewcommand{\theequation}{A2.\arabic{equation}}
\setcounter{equation}{0}
We normalise the areas by choosing the simplest form of the
Schwarz--Cristoffel transformations that map the disc--like
domains to
the upper half complex plane. Let $z$ be the variable in the domain,
and $\omega$ the variable in the upper half plane. Then we
have:
\begin{equation}
\label{SC}
{{d\omega}\over{dz}}
= \phi_D\prod_j (\omega-\omega_j)^{1-\alpha_j}
\end{equation}
The vertices of the domain map to
the branch points $\omega_j$, which are determined up to a rational
fractional transformation of the upper half plane.
This redundancy may be eliminated (up to permutations)
by fixing the positions
of the images of three vertices.
An overall
phase factor $\phi_D$ is determined by the orientation of
the domain $D$ in the $z$--plane.
For the simplest case, triangles, the three branch points may
be put in $0$, $1$ and $\infty$ in the $\omega$--plane, and the
side of the triangle
between the vertices that are mapped to $0$ and $1$ can be
placed parallel to the real axis in the $z$--plane.
We then have the familiar formula:
\begin{equation}
\label{triangleSC}
{{d\omega}\over{dz}}
= \omega^{1-\alpha_0}(1-\omega)^{1-\alpha_1}
\end{equation}

A variation of the shape of the domain,
brings a simultaneous variation of all the
angles, and of the positions of the branch points:
the expression for the variation hence concretely depends
on the parametrization chosen of the branch points in
terms of the opening angles and the side lengths.
Let us first consider the behaviour close to a branch point, say
$\omega_0$. By a linear change of variables,
$\tilde\omega = \omega -\omega_0$ before the variation,
and
$\tilde\omega = \omega -\omega_0'$ after the variation,
we can consider
the branch point $\omega_0$ to lie at the origin, and
not to move under the variation.
Taking a small corner around the inverse image of the branch point
(in the $z$--plane), and ignoring terms that vary slowly over the
corner, we have
\begin{eqnarray}
\label{varycorner}
2\delta\sigma_{\hbox{corner}}
= \log |{{dz'}\over{dz}}|^2 =   \log |{{dz'}\over{d\tilde\omega}}|^2
- \log |{{dz}\over{d\tilde\omega}}|^2 = \nonumber \\
\delta\alpha_0\log |\alpha_0 z|^{2\over{\alpha_0}}
+ \sum_{j\neq 0} \delta\alpha_j\log |\omega_j-\omega_0|^2
- \sum_{j\neq 0} (1-\alpha_j)
[{{\delta(\omega_j-\omega_0)}\over{\omega_j-\omega_0}} + \hbox{c.c}]
\end{eqnarray}
In the bulk (i.e. away from the corners), we do not shift the
$\omega$--coordinates, and have instead:
\begin{eqnarray}
\label{varybulk}
2\delta\sigma_{\hbox{bulk}}
= \log |{{dz'}\over{dz}}|^2 =   \log |{{dz'}\over{d\omega}}|^2
- \log |{{dz}\over{d\omega}}|^2 = \nonumber \\
 \sum_{j} \delta\alpha_j\log |\omega-\omega_j|^2
- \sum_{j} (1-\alpha_j)
[{{\delta\omega_j}\over{\omega-\omega_j}} + \hbox{c.c}]
\end{eqnarray}
However, by Appendix 3, in piece--wise flat domains, the only
bulk contribution will arise from the normal derivative
at the boundary of \ref{varybulk}, so let us compute that
(in the $\omega$-plane, that is, at the real axis):
\begin{eqnarray}
\label{normderiv}
2\hat n\cdot\nabla \delta\sigma_{\hbox{bulk}}=
-2\sum_{j} \delta\alpha_j{{\Im\omega_j}\over{ |\omega-\omega_j|^2}}
+2 \sum_{j} (1-\alpha_j)
\Im[{{\delta\omega_j}\over{(\omega-\omega_j)^2}}]
\end{eqnarray}
If $\omega_j$ is on the real axis, \ref{normderiv} vanishes
identically. If $\omega_j$ lies in the upper complex plane
(an interior corner) we are to integrate the normal
derivative along the boundary. The second sum in
\ref{normderiv} then clearly gives zero, while the first
integrates to
\begin{eqnarray}
\label{normintegrate}
\int_R 2\hat n\cdot\nabla \delta\sigma_{\hbox{bulk}} ds=
-4\pi\sum_{\Im\omega_j > 0} \delta\alpha_j
\end{eqnarray}

For simplices with disc topology, all the contributions
except the rather simple result \ref{normintegrate}
therefore
come from the corners, and
for polygons, there are no interior corners, and
\ref{varycorner} gives the only relevant terms.
We have not been able to put \ref{varycorner} in a form
which is manifestly a total variation, but in all the
special cases where we checked it, it turns out to
be so. In some cases
it is possible to choose the parametrization of a family
of polygons
such that the corners on the boundary do not move: in this
case all the contributions are from the first two terms
in \ref{varycorner}.
For triangles we have the final simplification in that we can
use the parametrization
(\ref{triangleSC}), and then the only contributions are from the
first term in \ref{varycorner}.

\section* {A3. Contribution of a straight boundary}
\label{s:app3}
\renewcommand{\theequation}{A3.\arabic{equation}}
\setcounter{equation}{0}
On piece--wise flat surfaces, away from the corners,
we can always approximate the heat kernel with the
free space heat kernel, or the heat kernel close to a
straight boundary. The error term ($K_D - \tilde K_D$
in \ref{intdivided}) is exponentially small in ${1\over t}$.
In free space the diagonal elemeent of the heat kernel
is $K_D(x,x,t) = {1\over{4\pi t}}$. That leads to a contribution
to $Z(x,x,s)$ from $\tilde K_D$
in \ref{intdivided}, which has
a singularity at $s=1$ with residue ${1\over{4\pi}}$,
but which is regular at lower $s$.
Inside the piece--wise flat surfaces
there are therefore no contributions to $Z(0)$,
or to the variation of $Z'(0)$.

The heat kernel in the half plane with Dirichlet conditions
on the $x$--axis is expressed as the difference between a
``direct'' and a ``reflected'' term:
\begin{equation}
K((x_1,y_1),(x_2,y_2),t) = {1\over{4\pi t}}
(\exp(-{{(x_1-x_2)^2+(y_1-y_2)^2}\over{4 t} })-
\exp(-{{(x_1-x_2)^2+(y_1+y_2)^2}\over {4t} }))
\end{equation}
The diagonal elements are then
\begin{equation}
K((x,y),(x,y),t) = {1\over{4\pi t}}
(1-\exp(-{{y^2}\over t }))
\end{equation}
We compute the Mellin transform multiplied
with a convergence factor $\exp (-\mu t)$, and $s$
between ${1\over 2}$ and $1$:
\begin{equation}
{1\over {\Gamma(s) 4\pi}}\int_0^{\infty} dt t^{s-1}
{{ 1 - \exp(-{{y^2}\over t }})\over{t}} e^{-\mu t}
= - {{\Gamma(1-s)}\over{4\pi\Gamma(s)}} ({{y^2}})^{s-1} +
{\cal O} (\mu^{2-s},\mu)
\end{equation}
To investigate the contributions to $Z(s)$ we may integrate $y$
over a strip along the boundary with width $\delta$ and length $L$.
We then obtain:
\begin{eqnarray}
- L{{\Gamma(1-s)}\over{4\pi\Gamma(s)}} {{\delta^{2s-1}}\over{2s-1}}
\end{eqnarray}
When we continue to lower values of $s$ we find
a pole at $s={1\over 2}$ with residue
$-{L\over{8\pi}}$
which implies that the short term asymptotics of the heat kernel
has one term which is a line density at
the boundary, with coefficient $-t^{-{1\over 2}}{1\over{8\pi^{1\over 2}}}$.

At the origin, the contribution goes as $s$, hence there is no
contribution to $Z_D(0)$ from a straight edge. We obtain contributions
to $Z_D'(0)$ as
\begin{eqnarray}
\label{boundary}
{L\over{4\pi\delta}}
\end{eqnarray}
The interpretation of \ref{boundary} is, that apart from the
finite and regular terms in $Z'(x,x,0)$, there is a distribution,
which is defined by the analytic continuation of $y^{2s-2}$.
If we integrate over strips parallel to the boundary, the
terms as in \ref{boundary} would cancel pairwise, and we would
be left with the outer boundary term. Hence we conclude that
a straight boundary gives a contribution to $Z_D'(x,x,0)$
which is no more singular than a line density, and
hence gives a contribution proportional to boundary length.
In general
one could integrate $Z_D((x,y),(x,y),s)$ against a smooth
test--function $f(x,y)$,
with $s$ in the interval $[{1\over 2}, 1]$,
and then analytically continue to smaller $s$. Such a test--function
is the conformal variation $\delta\sigma(x,y)$. Assuming an expansion
normal to the boundary at $x$ as
$2\delta\sigma(x,y) = a_0(x) + y\cdot a_1(x) + \ldots$ we find
\begin{eqnarray}
\label{distr}
\int_0^L dx\int_0^{\delta} dy 2\delta\sigma(x,y) Z((x,y),(x,y),s) \nonumber \\
= \int_0^L dx\int_0^{\delta} dy
[ - {{\Gamma(1-s)}\over{4\pi\Gamma(s)}} ({{y^2}})^{s-1} ]
[ a_0(x) + y\cdot a_1(x) + \ldots]  \nonumber \\
=  - [{{\Gamma(1-s)}\over{4\pi\Gamma(s)}} {{{\delta}^{2s-1}}\over{2s-1}}
\int_0^L dx a_1(x)] -
[{{\Gamma(1-s)}\over{4\pi\Gamma(s)}} {{{\delta}^{2s}}\over{2s}}
\int_0^L dx a_1(x)] + \ldots
\end{eqnarray}
The contribution to the variation of $Z'_D(0)$ is the limit of
\ref{distr} as $s$ tend to zero, which may clearly be expressed in
the normal derivative at the boundary of the conformal variation:
\begin{eqnarray}
\delta Z'_{\hbox{straight boundary}}(0) = -{1\over{4\pi}}
\int_0^L \hat n \cdot  \nabla(\delta\sigma) ds
\end{eqnarray}

\section* {A4. The Sommerfeldt kernel}
\label{s:app4}
\renewcommand{\theequation}{A4.\arabic{equation}}
\setcounter{equation}{0}
A. Sommerfeldt in 1896 solved the problem of diffraction of light
by a perfectly conducting half--plane\cite{SOpt}. The solution
takes the form of a kernel periodic in the angle variable with periodicity
$4\pi$; the difference of one ``direct'' and one ``reflected'' wave
vanishes at $0$ and $2\pi$.

We will need the solution to the diffusion problem in a sector with
opening angle $\pi\alpha$, which is quite analogous, but for completeness
given here. (The solution of diffusion problem at an interior
corner with total angle $2\pi\alpha$ is obtained in the same way,
by keeping only the ``direct'' term.)
If the opening angle is of the form ${{\pi}\over n}$ the sector
can be refleced in its side $2n$ times to precisely cover $2\pi$, and the
solutions to both the diffusion problem and the diffraction problem  are
expressed by the method of images. Sommerfeldts solution is a substitute
when the reflections do not make up a full turn, and is given by
a certain finite number of image charges, and a correction term.
It has the following integral representation:
\begin{equation}
K_S(r,\phi ; r',\phi' ; t) = {1\over {4\pi t}}
\exp ( - { {r^2+{r'}^2}\over{4t} })
 [ \nu_{\alpha}({{rr'}\over{2t}},\phi-\phi')
- \nu_{\alpha}({{rr'}\over{2t}},\phi+\phi')]
\end{equation}
where the important part is
\begin{equation}
\nu_{\alpha}(a,\phi) = {1\over {2\pi\alpha}}
\int_{A+B} \exp ( a\cos \delta ) {{d\delta}\over
{1-e^{-{{i(\delta +\phi)}\over\alpha}}}}
\end{equation}
$A$ and $B$ are paths in the plane of complex $\delta$ that go asymptotically
from $\pi + i\cdot\infty$ to $-\pi + i\cdot\infty$, and  $-\pi - i\cdot\infty$
to $\pi - i\cdot\infty$, respectively.
Essentially this is a superposition of free heat kernels between $ x$ and
$y'$, where $|y'| = r'$ and
$| x-y'|^2 = r^2+r'^2 - 2rr'\cos(\delta )$. In the bands
${{(4n+1)\pi}\over 2} < |Re(\delta )| < {{(4n+3)\pi}\over 2}$
the contour integral can be taken to infinity since
$\Re(| x-y'|^2 ) \to \infty$,
but not in between.

$K_S$ satisfies the heat equation because it is superposition of free heat
kernels, it is symmetric in $(x,y)$
and periodic in $\phi$ and $\phi'$ with period $2\pi\alpha$
by construction of $\nu_{\alpha}$,
and it vanishes at the boundaries of the sector because it is the
difference between a direct and a reflected term.
Furthermore, away from the imaginary axis it is analytic in $\alpha$.

By deforming $A$ and $B$ into the straight lines $\pi + iy$, $-\pi+iy$ and
$[-\pi,\pi]$, $\nu_{\alpha}$ can be written as
\begin{eqnarray}
\nu_{\alpha}(a,\phi) = \sum_{k: -1<2\alpha k - \phi/\pi < 1}
\exp(a\cos(2\pi\alpha k - \phi )\nonumber \\
- {{\sin(\pi/\alpha )}\over {2\pi\alpha}}
\int_{-\infty}^{\infty}{{\exp(-a\cosh y) }\over{\cosh (y/\alpha - i\phi/\alpha)
-\cos \pi/\alpha }} dy
\end{eqnarray}
{}From the image charges of the ``direct'' term, it
follows that the normalization of the kernel is correct.
If for some $k$ an image charge wanders
through the line $\pm \pi$,
we ought to take half the residue and the principal value of the integral
in the usual fashion, but then
$\alpha$ equals ${{\pi}\over n}$,
and the prefactor of the integral is zero: that is;
we have just a solution by images.

\section* {A5. A useful integral}
\label{s:app5}
\renewcommand{\theequation}{A5.\arabic{equation}}
\setcounter{equation}{0}

In this appendix we discuss an elementary but useful integral. We would
like to compute a contribution to $Z_D(0)$ or $Z_D'(0)$ by
substituting the true (unknown) heat kernel in a compact
domain, with essentially something
like a heat kernel in the full plane. This includes the case with
a flat boundary, where the heat kernel can be written as the difference
between a free heat kernel and an image charge reflected over the
boundary, and also the case in a sector, since the
Sommerfeldt heat kernel can be written as a linear
superposition of free heat kernels.
The first problem is that the eigenvalues in an infinite domain
accumulate to zero; the Mellin transforms are therefore not
convergent for any parameter $s$. We are however interested
in Mellin transforms of heat kernels in compact domains, which are
asymptotic to free heat kernels for short times, but fall of
exponentially for large times. Furthermore, we are only
interested in the singular contribution, so the problem can be
resolved by limiting the integral in the Mellin transform to the interval
$[0,1]$. In practice such an integral is not analytically tractable. Instead
we can introduce a convergence factor $e^{-\mu t}$, including
which the Mellin transform converges for $Re(s) > 1$,
and look for the singular part in
the $\mu$--independent part of the result.

We therefore consider
\begin{equation}
\int_0^{\infty} t^{s-1} e^{-{a\over t}} e^{-\mu t} dt
=  2({a\over{\mu}})^{s\over 2} K_s((4a\mu )^{1\over 2})
= \Gamma (s) \mu^{-s} + \Gamma(-s) ({a})^{s} + {\cal O} (\mu^{1-s},\mu)
\end{equation}
by the power series expansion of the modified Bessel function.

Two variants, with $s$ around the origin, are
\begin{equation}
{1\over{\Gamma(s)}}\int_0^{\infty} t^{s-1}(1 - e^{-{a\over t}}) e^{-\mu t} dt
= -{{\Gamma(-s)}\over {\Gamma(s)}} ({a})^{s} + {\cal O} (\mu^{1-s},\mu)
\end{equation}
and
\begin{equation}
{1\over{\Gamma(s)}}\int_0^{\infty} t^{s-1}{{e^{-{a\over t}}}\over t} e^{-\mu t}
dt
= {1\over{(s-1)}}\mu^{1-s} + {{\Gamma(1-s)}\over {\Gamma(s)}}({a})^{s-1} +
{\cal O} (\mu^{2-s},\mu)
\end{equation}
As $\mu$ tends to zero, in the first case, the value at zero is 1,
and the derivative is
$2\gamma + \log a $ by the expansion
$\Gamma (s) \sim {1\over s} - \gamma + \ldots$
In the second case the value at zero vanishes, but the derivative is
${1\over a}$.
\section* {A6. The singular corner contribution}
\label{s:app6}
\renewcommand{\theequation}{A6.\arabic{equation}}
\setcounter{equation}{0}
We will here compute the contributions to $Z_D(0)$
and $\delta Z_D'(0)$ from a corner.
For $Z_D(0)$ we are by Appendix 2
and Appendix 4 looking
for the limit as $s$ tends to zero of
\begin{equation}
\label{z0contribution}
\int_{\hbox{corner}} d^2 x Z(x,x,s)
\end{equation}
where the zeta function density is computed from
the Sommerfeldt heat kernel, and the corner is delimited
by a cut-off radius $r_0$.

For $\delta Z_D'(0)$ we use the variational
formula (\ref{varyPERS}), and look
for the term finite in $s$ as $s$ tends
to zero of
\begin{equation}
\label{zprim0contribution}
\int_{\hbox{corner}} d^2 x  [2\delta\sigma(x)Z(x,x,s)]
\end{equation}
An inspection of \ref{varycorner} shows that the conformal
factor close to the corner contains two types of terms, one
which is proportional to the logarithm of the distance to
the tip of the corner, and one which is constant over the corner.

We choose in the following to keep the constant factors
in first term in \ref{varycorner}, so that what we compute
will directly be the stricly local
contribution to $\delta Z_D'(0)$, which can then be given in
integrated form. The contributions of the last two terms
in \ref{varycorner} will be proportional to the
contribution to $Z_D(0)$, and are easily included by
comparison.

Our computations therefore boil down to \ref{z0contribution}
and
\begin{eqnarray}
\label{intonconstant}
\int_0^{\pi\alpha} \int_0^{r_0}r d\phi dr [{{\delta\alpha}\over{\alpha}}
 2\log\alpha Z_{\alpha}(x,x,s) ] \\
\label{intonlog}
\int_0^{\pi\alpha} \int_0^{r_0}r d\phi dr [{{\delta\alpha}\over{\alpha}}
 \log r^2 Z_\alpha (x,x,s) ]
\end{eqnarray}
where we can directly integrate over
$\phi$ since our test functions (a constant and $\log r^2$) do not
depend on $\phi$.

It will turn out that the reflected term in the Sommerfeldt
kernel only contributes to $\delta Z_D'(0)$, and
only by a simple term.
We can therefore in parallel compute the contributions to a corner
at the boundary with opening angle $\pi\alpha$, and a corner in the interior
with opening angle $2\pi\alpha$, which almost only differ by a factor
of two.
By analyticity in $\alpha$ we can choose  the convenient range
${1\over 2} < \alpha < 1$.

\subsection*{The image charges}

The charge from the direct term, and the
image charges from the reflected
term give contributions to the Mellin
transform, integrated over the area of the corner, as
\begin{equation}
\label{refintegral}
{1\over {4\pi t}}
\int_0^{r_0} r dr \int_0^{\pi\alpha} d\phi [
1 -
\sum_{k: -1<2\alpha k - 2\phi/\pi < 1}
\exp(-{{r^2}\over{2t}}(1-\cos(2\pi\alpha k - 2\phi )))]
\end{equation}
The direct term charge can be neglected, as along a straight boundary.
With ${1\over 2} < \alpha < 1$ the only relevant values of $k$
in the reflected term are
$0$ and $1$.
For $k=0$ the integral over $\phi$ is limited between 0
and ${{\pi}\over 2}$, while
for $k=1$ it goes between $\pi\alpha - {{\pi}\over 2}$ and $\pi\alpha$. With
$y=r\sin(\phi )$ in the first case, and  $y=r\sin(\phi - \pi\alpha )$
in the second, we find with $s$ between $1\over 2$ and $1$:
\begin{eqnarray}
\int_0^{{\pi}\over 2} d\phi r Z_{\alpha}((r,\phi),(r,\phi),s) &=&
- {{\Gamma(1-s)}\over{4\pi\Gamma(s)}}r^{2s-1}
 2\int d\phi \sin^{2s-2}\phi \nonumber \\
&=&-{{\Gamma(1-s)r^{2s-1}
	\Gamma({1\over 2})\Gamma(s-{1\over 2})}
	\over
	{4\pi\Gamma(s)\Gamma(s)}}
\end{eqnarray}
When we integrate up to a radius $r_0$
and continue to lower $s$ we find, as we should, a pole at $s={1\over 2}$
with residue $-{{2r_0}\over{8\pi}}$.

To investigate the contributions close to zero, we use
the formulae
\begin{eqnarray}
\label{rintegrations}
\int_0^{r_0} dr r^{2s-1} = {{r_0^{2s}}\over{2s}}
 = {1\over {2s}} + {1\over 2}\log r_0^{2} + {\cal O}(s) \nonumber \\
\int_0^{r_0} dr r^{2s-1}\log r^2 =
{{\partial}\over{\partial s}} [{{r_0^{2s}}\over{2s}}]
 = -{1\over {2s^2}} + {1\over 4}(\log r_0^{2})^2 + {\cal O}(s) \nonumber
\end{eqnarray}
where we understand that we first integrate over $r$, and then
analytically continue to the neighbourhood of the origin.

The contribution to $\delta Z'_{\alpha}(0)$ from the reflected image
charges (hence only for boundary corners) is thus
\begin{eqnarray}
\label{imagecontr}
 - {{\delta\alpha}\over{4\alpha}}
\end{eqnarray}

\subsection*{The reflected term: the integral}
The integral part of the reflected term comes from
\begin{equation}
{{\sin \pi/\alpha}\over {8\pi^2\alpha t}}
\int_0^{\pi\alpha} r d\phi \int_{-\infty}^{\infty}
{{\exp (-{{r^2 (1 + \cosh y )}\over{2t}} )}\over
{\cosh  (y/\alpha - 2i\phi/\alpha ) - \cos \pi/\alpha }} dy ,
\end{equation}
where we have integrated
over the angle variable $\phi$.
The $\phi$--integration forms part of closed contour from $-i\infty$ to
the origin, along the real axis to $\pi\alpha$, and down parallel to the
imaginary axis to $\pi\alpha - i\infty$. The integrals over the two lines
cancel. If $y>0$ there are two poles inside the contour, but their
residues cancel. The angle integral is thus zero.

\subsection*{The direct term: the integral}
We consider
\begin{eqnarray}
\label{direct1}
-{{\sin \pi/\alpha}\over {8\pi^2\alpha t}}
\int_0^{\pi\alpha} r d\phi\int_{-\infty}^{\infty}
{{\exp (-{{r^2 (1 + \cosh y )}\over{2t}} )}\over
{\cosh y/\alpha - \cos \pi/\alpha }} dy
\end{eqnarray}
Following Appendix 2, the
Mellin transform of \ref{direct1},
performed termwise inside the integral over $y$,
gives
\begin{equation}
-r^{2s-1} {{\Gamma (1-s)\sin  (\pi/\alpha) }\over {\Gamma (s) 8\pi }}
\int_{-\infty}^{\infty}
{1 \over
{(\cosh y/\alpha - \cos \pi/\alpha )(\cosh^2 y/2)^{1-s}}} dy
\end{equation}
The integral is most tractable if considered as a correlation
integral between the two factors in the denominator, then
by Percival's formula\footnote{This transformation brings us closer to
the Lebedev--Kantorovich tranform of D. Ray, that was
used by McKean and Singer in \cite{McKean}.}
we have
\begin{equation}
r^{2s-1}{{\Gamma (1-s) \alpha }\over {\Gamma (s) 8\pi }}
2^{2-2s}\int_{-\infty}^{\infty} B (1-s +iy,1-s-iy) [
{{\sinh \pi y \cosh \pi\alpha y - \sinh \pi\alpha y \cosh \pi y}\over
{\sinh \alpha\pi y }}] dy
\end{equation}
This expression can be written more compactly as
\begin{equation}
\label{Zscompact}
r^{2s-1}{{\Gamma (1-s) \alpha 2^{2-2s} }\over {\Gamma (s) 8\pi }}
A(s)
\end{equation}
where the numerical values are determined by the
integral
\begin{equation}
\label{Asdef}
A(s) = \int_{-\infty}^{\infty} B (1-s +iy,1-s-iy) [
{{\sinh \pi y \cosh \pi\alpha y - \sinh \pi\alpha y \cosh \pi y}\over
{\sinh \alpha\pi y }}] dy
\end{equation}

Using the identity
\begin{equation}
B (1+iy,1-iy) = {{\pi y}\over
{\sinh \pi y }}
\end{equation}

we have
\begin{equation}
A(0) = \int_{-\infty}^{\infty} [\pi y \coth \alpha\pi y -
\pi y \coth \pi y] dy \\
\end{equation}
and
\begin{equation}
A'(0) = \int_{-\infty}^{\infty}(2\psi(2)-\psi(1+iy) -\psi(1-iy)) [\pi y \coth
\alpha\pi y -\pi y \coth \pi y] dy
\end{equation}

where $\psi(x+1) = -\gamma +\sum_{n=1}^{\infty} ({1\over n} - {1\over {n+x}})$.

\subsection*{Contribution to $Z_D(0)$}
We integrate
\ref{Zscompact}
over the radial variable:
\begin{equation}
{{r_0^{2s}}\over{2s}}[{{s\alpha}\over{2\pi}} A(0) + {\cal O} (s^2)]
\end{equation}

$A(0)$ can be computed by introducing a factor $e^{i\epsilon y}$, and closing
the integral over $y$ in the upper half plane.
The residues contribute
\begin{eqnarray}
-2\pi \sum_{n=1}^{\infty} n({1\over{\alpha^2}}e^{-n\epsilon/\alpha} -
e^{-n\epsilon}) \nonumber \\
= -{\pi\over 2} ( {1\over {\alpha^2\sinh^2 (\epsilon/2\alpha )}}
- {1\over {\sinh^2 (\epsilon/2)}})
\end{eqnarray}
which, in the limit as $\epsilon \to 0$, becomes ${\pi\over 6}(
{1\over {\alpha^2}}
- 1)$

The value of $Z_{\alpha}(0)$ is therefore
\begin{eqnarray}
\label{z0magic}
{1\over 24}({1\over {\alpha}} - \alpha)
\end{eqnarray}
a result first derived by
McKean and Singer\cite{McKean}.

\subsection*{Contribution to $\delta Z_D'(0)$}

Using again \ref{rintegrations} and the expansion of $A(s)$
and the prefactors
at the origin, we obtain:
\begin{eqnarray}
\label{sexpansion}
[{{\delta\alpha}\over{\alpha}}]
[s{{\alpha}\over{2\pi}} A(0) +
{{s^2\alpha}\over{2\pi}}(A'(0) + A(0)(2\gamma  -2\log 2))]
[- {1\over {2s^2}} + { { \log \alpha }\over s}] + \hbox{h.o.t}
\end{eqnarray}
We have a pole in $s$:
\begin{eqnarray}
-{1\over {s}} {{\delta\alpha}\over{\alpha}}{\alpha\over{4\pi}} A(0)
\end{eqnarray}
Since ${{\alpha}\over{4\pi}} A(0)$ is the contribution to $Z_D(0)$
from the corner, we can write this variation as
\begin{eqnarray}
\label{modvaryZ}
{1\over s} {{\delta\alpha}\over{24}}(-{1\over{\alpha^2}}+1)
= {1\over s} \delta [{1\over{24}}({1\over{\alpha}}+\alpha)]
\end{eqnarray}
Let us work out that this variation is only seemingly
in contradiction
with the known expression of $Z_{\alpha}(0)$ in \ref{z0magic}.
The variation of $Z_D(x,x,0)$
is of course a point mass at the corner, with weight
${{\delta\alpha}\over{24}}(-{1\over{\alpha^2}}-1)$.
The variational forms (\ref{varyT}) and (\ref{smallvaryT})
are however for the variation of $Z_D(s)$, not for the
zeta function density. The variations we consider go between
different simplices with disc--like topology, and therefore leave
the sum $\sum_{\hbox{i.c.}}(2-2\alpha_j) + \sum_{\hbox{b.c.}}(1-\alpha_j)$
invariant ({\it see} equation~\ref{invariant}).
The analytical expressions of the variation are then
only determined up to terms linear in the variations of the angles,
which is the discrepancy we have in \ref{modvaryZ}.

The contribution finite in $s$ in \ref{sexpansion} is
\begin{eqnarray}
\delta\alpha
[{{ A(0)}\over{4\pi}}(2\log\alpha-2\gamma + 2\log 2)
-{{ A'(0)}\over{4\pi}}]
\end{eqnarray}
We compute $A'(0)$ by
introducing a convergence factor $e^{i\epsilon y}$.
Since $\psi(\bar z ) = \bar \psi (z)$, it is enough to consider
\begin{equation}
C_n = \int_{-\infty}^{\infty}({1\over n} - {1\over {n-iy}}) [\pi y \coth
\alpha\pi y -\pi y \coth \pi y] dy
\end{equation}
in terms of which $A'(0) = 2 A(0) - 2\Re\sum_{n=1}^{\infty}C_n$.
The two terms in $C_n$ have residues in the upper half plane, and sum to
\begin{eqnarray}
&\, &-{{\pi}\over {2n}} ( {1\over {\alpha^2\sinh^2 (\epsilon/2\alpha )}}
- {1\over {\sinh^2 (\epsilon/2)}}) \nonumber \\
&\, &+ 2\pi \sum_{m=1}^{\infty}
m({1\over{\alpha^2(n+m/\alpha)}}e^{-m\epsilon/\alpha} -
{1\over{(n+m)}}e^{-m\epsilon}) \nonumber \\
&=& - {{\pi}\over {6n}} ( {1\over {\alpha^2}} - 1)
 + \pi(1-{1\over{\alpha}})  \nonumber \\
&\, & -2\pi ne^{\epsilon n}\int_{\epsilon}^{\infty}
e^{-\mu n} [{1\over{\alpha(e^{\mu/\alpha} -1) }}- {1\over{(e^{\mu} -1) }}] d\mu
+ {\cal O} (\epsilon )
\end{eqnarray}

With some care the sum in $n$ is:
\begin{eqnarray}
\label{cruzint}
&\, & \sum_{n=1}^{\infty} C_n = -2\pi I(\alpha) = \nonumber \\
&\,&-2\pi
\int_{0}^{\infty}
{1\over{e^{\mu} -1}}
[{1\over {4\sinh^2 (\mu/2)}} - {1\over {4\alpha^2\sinh^2 (\mu/2\alpha )}}
-{1\over {12}}({1\over{\alpha^2}}-1)] d\mu
\end{eqnarray}

Now we can collect all the strictly local contributions
to $Z_D'(0)$ and write
\begin{eqnarray}
\label{totalvary}
\delta Z_{\alpha}'(0) = -{{\delta\alpha}\over{4\alpha}} +
\delta\alpha [{1\over{12\alpha}}({1\over{\alpha}} - \alpha)
(\log\alpha-\gamma + \log 2 -1) - I(\alpha)]
\end{eqnarray}
Here we have taken a boundary corner, and hence included the
contribution from the reflected image charges.

It is clearly of interest to give \ref{totalvary} in
integrated form, whence we introduce the primitive function
of $I(\alpha)$:
\begin{eqnarray}
\label{Jdef1}
J(\alpha) =
\int_{0}^{\infty}
{1\over{e^{\mu} -1}}
[{1\over {2\mu}}\coth({{\mu}\over{2\alpha}}) -
{{\alpha}\over {4\sinh^2 ({{\mu}\over 2})}}
-{1\over {12}}({1\over{\alpha}}+\alpha)] d\mu \\ \nonumber
I(\alpha) = -{d\over{d\alpha}} J(\alpha)
\end{eqnarray}
We know from Appendix 3 that the contribution to $Z_D'(0)$
({\it note:} not the variation) of a straight boundary
is proportional to the boundary length. The integrated form
of \ref{totalvary} should therefore be zero at $\alpha=1$.
As in the variation of $Z_D(0)$, we should in addition
expect undetermined linear terms in the angles.

A rather long derivation, to be given in Appendix~8 as we
investigate the special values of $\alpha={1\over n}$, shows that
we may integrate \ref{totalvary}, compare with one of
the integrable cases, include the integration constant in
the contribution from the corners, and simplify to:
\begin{eqnarray}
\label{tibc}
Z'_{\alpha}(0) = {1\over{12}}({1\over{\alpha}} -\alpha)(\gamma -\log 2)
-{1\over{12}}({1\over{\alpha}} + 3 +\alpha)\log\alpha +\tilde J(\alpha)
\end{eqnarray}
where the last term has the more symmetric integral representation
\begin{eqnarray}
\label{Jdef2}
\tilde J(\alpha) =
\int_{0}^{\infty}
{1\over{e^{x} -1}}
[{1\over {2x}}(\coth({x\over{2\alpha}}) -
\alpha \coth({x\over{2}}))
-{1\over {12}}({1\over{\alpha}}-\alpha)] dx
\end{eqnarray}

\subsection*{Difference at interior corners}
The only change is that the angle--independent direct term
is integrated over twice as wide an angle, and the contribution
from the reflected image charges are absent.
The contribution to $Z_D(0)$ is then
\begin{eqnarray}
Z_{\alpha}(0) = {1\over {12}}({1\over {\alpha}} - \alpha)
\end{eqnarray}
and the strictly local contributions to $Z_D'(0)$
\begin{eqnarray}
\label{tiic}
Z'_{\alpha}(0) = {1\over{6}}({1\over{\alpha}} -\alpha)(\gamma -\log 2)
-{1\over{6}}({1\over{\alpha}} + \alpha)\log\alpha +2\tilde J(\alpha)
\end{eqnarray}

\subsection*{Summary of corner contributions}
For simplices with the disc topology, the contributions to the
variation of the zeta function derivative are given by \ref{varycorner}
and a simple sum over the interior corners.
There are three terms in \ref{varycorner}, of which the first is
the only one that arises in triangular domains. It is written
in final variational form valid for corners
on the boundary in \ref{totalvary}, \ref{cruzint}, and
in integrated form in \ref{tibc}, \ref{Jdef2}.
The last two terms in \ref{varycorner} are of the same type
as \ref{intonconstant}, i.e. in variation form they will
give something proportional to the contribution to $Z_D(0)$
from the corner. We give those expressions here, under the
(slight) simplifying assumption that none of the branch point
in the $\omega$--plane lies at infinity:
\begin{eqnarray}
\label{varyextra}
\delta Z'_{\hbox{extra}} = \sum_{i,j}
Z_{\alpha_i}(0)[
\delta\alpha_j\log |\omega_j-\omega_i|^{-2}
-(1-\alpha_j)
[{{\delta(\omega_j-\omega_i)}\over{\omega_j-\omega_i}} + \hbox{c.c}]]
\end{eqnarray}
\section* {A7. Computation using the heat kernel}
\label{s:app7}
\renewcommand{\theequation}{A7.\arabic{equation}}
\setcounter{equation}{0}
This approach was used by Dowker\cite{Dowker} to compute the
contribution to $Z_D(0)$ from a corner.
We begin with the asymptotic formulae for the trace
of the heat kernel in a domain $D$:
\begin{eqnarray}
\label{tracec}
Tr_x [ K_D(x,x,\epsilon)] &\sim& {{c_1 }\over {\epsilon}} +
{{c_{1\over 2}  }\over {\epsilon^{1\over 2}}}
+Z_D(0) +
{\cal O}(\epsilon^{1\over 2})\\
\label{varytracec}
Tr_x[2\delta\sigma(x) K_D(x,x,\epsilon)] &\sim&
{{\delta[c_1] }\over {\epsilon}} +
{{2\delta[c_{1\over 2}]  }\over {\epsilon^{1\over 2}}}
-\log\epsilon\, \delta[Z_D(0)]
-\gamma\delta [Z_D(0)] \nonumber \\
&+& \delta[Z_D'(0)] +
{\cal O}(\epsilon^{1\over 2})
\end{eqnarray}
As in Appendix~6 we only need to consider the image charges
from the reflected term and the integral from the
direct term.

\subsection*{The direct term: the integral}
It is convenient to go back to the original form
of the Sommerfeldt kernel (including the line integrals
and the direct charge):
\begin{equation}
K_{\hbox{direct}}(r,\phi ; r,\phi ; t) = {1\over {8\pi^2\alpha t}}
\int_{A+B}
{{d\delta}\over
{1-e^{-{{i\delta}\over\alpha}}}} e^{-{{r^2}\over{2t}}(1-\cos \delta)}
\quad - (\hbox{the pole at $\delta=0$})
\end{equation}
We are to integrate the kernel, by itself or
multiplied by ${{\delta\alpha}\over{\alpha}}
\log \alpha^2r^2$, over the corner, and then
catch the finite piece as $t$ tends to zero.
The integration contour is chosen such that the prefactor
of $r^2$ in the exponent always has negative real part.
We can therefore extend
the integral over all $r$
(as in Appendix~5), without introducing significant errors.
For the contribution to $Z_D(0)$ we then find
\begin{eqnarray}
\int_0^{\pi\alpha} r d\phi
\int_0^{r_0} K_{\hbox{direct}}(x,x,t)
= -{1\over{8\pi}}\int_{A'+B'}
{{d\delta}\over
{1-e^{-{{i\delta}\over\alpha}}}} {1\over{1-\cos \delta}}
 + {\cal O}(e^{-C\cdot {{r_0^2}\over t}}
\end{eqnarray}
where we have afterwards moved over the integration contour to the
straight lines parallel to the imaginary axis at $\pi$
and $-\pi$.
Except for an error exponentially small in ${1\over t}$,
we have only left a term independent of $t$, that will
give us what we want. By a change of variables,
$\delta\to -\delta$, the integration
contour $A'$ goes into $-B'$, and vice versa, and the second
factor in the integrand (${1\over{1-\cos \delta}}$) is unchanged.
It is therefore sufficient to keep in the first factor,
the part odd under this reflection, and we have
\begin{eqnarray}
 -{1\over{16\pi i}}\int_{A'+B'}
d\delta \cot ({{\delta}\over{2\alpha}})
{1\over{1-\cos \delta}}
\end{eqnarray}
This integral is now in fact convergent at infinity, so we can close
it, and evaluate it by computing the residue of the pole at
the origin. Hence
\begin{eqnarray}
\int_0^{\pi\alpha} r d\phi
\int_0^{r_0} dr K_{\hbox{direct}}(x,x,t)
&=& -{1\over{8\pi}}2\pi i \hbox{Res}[
\cot ({{\delta}\over{2\alpha}})
{1\over{1-\cos \delta}}]_{\delta=0}\nonumber \\
&=& {1\over{24}}({1\over{\alpha}} - \alpha)
\end{eqnarray}
which is, of course, the right answer.

To compute the variation of $Z_D'(0)$, it is again convenient to
extend the integral over
the radius to infinity. We then find
\begin{eqnarray}
\int_0^{\pi\alpha} d\phi r
\int_0^{\infty}
{1\over{\alpha}}\log (\alpha^2 r^2) e^{-{{r^2}\over{2t}}(1-\cos\delta)}
= {{\pi t}\over{1-\cos\delta}}(-\gamma+\log {{2\alpha^2 t}
\over{1-\cos\delta}})
\end{eqnarray}
(we do not write out the variational factor $\delta\alpha$,
except where necessary),
and hence
\begin{eqnarray}
\label{smalltexpr}
\int_0^{\pi\alpha} r d\phi
\int_0^{r_0}dr [2\delta\sigma(x) K_{\hbox{direct}}(x,x,t)] =\nonumber \\
 -{1\over{8\pi\alpha}}\int_{A'+B'}
{{d\delta}\over
{1-e^{-{{i\delta}\over\alpha}}}} {1\over{1-\cos \delta}}
(-\gamma+\log {{2\alpha^2 t}\over{1-\cos\delta}})
\end{eqnarray}
Pulling out the term divergent as $\log t$ we have
\begin{eqnarray}
\log t {{\delta\alpha}\over{\alpha}}
{1\over{24}}({1\over{\alpha}} - \alpha)
\end{eqnarray}
which is equal to $-\log t\, \delta[Z_{\alpha}(0)]$, up to the
term linear in the the angle.

Taking into account that the finite part of \ref{smalltexpr}
is $\delta[Z_{\alpha}'(0)] - \gamma \delta[Z_{\alpha}(0)]$ we have
\begin{eqnarray}
\delta[Z_{\alpha}'(0)] = \delta[Z_{\alpha}(0)](-\log (2\alpha^2) +2\gamma)
\nonumber \\
- {{\delta\alpha}\over{8\pi\alpha}}\int_{A'+B'}
{{d\delta}\over
{1-e^{-{{i\delta}\over\alpha}}}} {1\over{1-\cos \delta}}
\log {1\over{1-\cos\delta}}
\end{eqnarray}
We choose the branch of $\log {1\over{1-\cos\delta}}$ symmetric under
$\delta\to -\delta$. Then we can again substitute for
${1\over{1-e^{-{{i\delta}\over\alpha}}}}$ its anti-symmetric
part,
and the integral is convergent at infinity.
Essentially $\log {1\over{1-\cos\delta}}$ is a function of
$\delta^2$, so it can be chosen to have a branch cut along the
negative real axis in the $\delta^2$--plane. That translates
to two branch cuts in the $\delta$--plane, along the positive
and negative real axis. Evenness under $\delta\to -\delta$
then determines the phase choice in the left half plane.
The best we can do now is to
pull both integration
contours in to the imaginary axis and integrate
over the branch cuts. In addition we have a singularity
at the origin. We separate out a small circle around the origin
with radius $\epsilon$, and an integral along the branch cuts from
$\pm i\epsilon$ to $\pm i\infty$. The integrals along the branch cuts
give
\begin{eqnarray}
{1\over{4\alpha}}
\int_{i\epsilon}^{i\infty}
d\delta \cot({{\delta}\over{2\alpha}}) {1\over{1-\cos \delta}} \nonumber \\
= {1\over{4\alpha}}
-{1\over{\epsilon^2}} - {1\over{12}}({1\over{\alpha}} - \alpha)
+ {1\over{8\alpha^2}} \int_{\epsilon}^{\infty}
dy \coth({y\over{2}}) {1\over{\sinh^2({y\over{2\alpha}})}}
\end{eqnarray}

Close to the origin we estimate ($\delta = \epsilon e^{i\theta}$):
\begin{eqnarray}
\log {1\over{1-\cos\delta}} &=& \log 2 - 2\log\epsilon
-2i\theta + {1\over{12}}\delta^2 + \ldots
\qquad\hbox{$\Re\delta>0$} \nonumber \\
&=& \log 2 - 2\log\epsilon -2i\theta +2i\pi + {1\over{12}}\delta^2
+ \ldots
\qquad\hbox{$\Re\delta<0$} \nonumber
\end{eqnarray}
which eventually gives
\begin{eqnarray}
{1\over{24}}({1\over{\alpha^2}} -1) (\log 2 -2\log\epsilon)
- {1\over{24}} + {1\over{2\epsilon^2}}
\end{eqnarray}
Collecting all terms from the direct integral we have
\begin{eqnarray}
\delta [Z_{\alpha}'(0)]_{\hbox{direct integral}}
&=& \delta\alpha[ {1\over{24}}({1\over{\alpha^2}} -1)
(-2\gamma+2\log(2\alpha) -2\log\epsilon)\nonumber \\
&\, & - {1\over{2\epsilon^2}} +{1\over{4\alpha}}
-{1\over{12}}({1\over{\alpha^2}} + 1) \nonumber \\
&\, & + {1\over{8\alpha^2}} \int_{\epsilon}^{\infty}
dy \coth({y\over{2}}) {1\over{\sinh^2({y\over{2\alpha}})}}]
\end{eqnarray}
\subsection*{The image charges}
We should compute the integral
\begin{equation}
\delta [Z_{\alpha}'(0)]_{\hbox{image charges}}
= -\delta\alpha[ \int_0^{{\pi}\over 2} r d\phi
\int_0^{r_0} dr {1\over{\pi\alpha t}}
\log(\alpha r) e^{-{{r^2}\over{2t}}(1-\cos 2\phi)}]
\end{equation}
The problem is that we have to separate out a $r_0$-dependent piece
diverging as $t^{-{1\over 2}}$, before we can have the finite value
at the origin, and as it stands it is not possible to extend
the integral over $r$ to infinity. If the integral would have been
over a small square, we could have separated in $x$ and $y$,
and then the integral would be much easier. Let us first see that the
segment of the square outside the
quadrant is unimportant, and then compute the integral over
the square.
The difference is
\begin{equation}
- \int_0^{r_0} dy \int_{\sqrt{r_0^2-y^2}}^{r_0} dx {1\over{2\pi\alpha t}}
\log(\alpha^2(x^2+y^2) ) e^{-{{y^2}\over t}}
\end{equation}
which is essentially
\begin{equation}
- \int_0^{r_0} dy {{y^2}\over{4\pi\alpha t r_0}}
[\log(\alpha^2 r_0^2) +{{y^2}\over{r_0^2}} + \ldots] e^{-{{y^2}\over t}}
\end{equation}
which only gives a contribution of order $t^{1\over 2}$.

The integral over the square is
\begin{equation}
- \int_0^{r_0} dx \int_0^{r_0} dy {1\over{2\pi\alpha t}}
\log(\alpha^2(x^2+y^2)  e^{-{{y^2}\over t}}
\end{equation}
After first integrating over $x$, we can extend the integral over $y$
to infinity and find:
\begin{equation}
- \int_0^{\infty} dx {1\over{2\pi\alpha t}}
(\log\alpha^2 +2r_0(\log r_0 -1) +\pi y + \ldots)e^{-{{y^2}\over t}}
\nonumber \\
= \hbox{Const.}\cdot t^{-{1\over 2}} - {1\over{4\alpha}}
\end{equation}

\subsection*{Summary}
The terms diverging with the cutoff $\epsilon$ can be brought
into the integral, and then give, after some algebra,
\begin{equation}
\delta [Z_{\alpha}'(0)]
= \delta\alpha[ -{1\over{12}}({1\over{\alpha^2}} -1)
(1+\gamma-\log 2\alpha) - {1\over{4\alpha}} + J'(\alpha)]
\end{equation}
which is the same result as in \ref{totalvary}.

\section* {A8. Corners with opening angle ${1\over n}$}
\label{s:app8}
\renewcommand{\theequation}{A8.\arabic{equation}}
\setcounter{equation}{0}

We proceed to simplify the the general expressions
of the contributions to the
regularized determinants, when the angles are of the form ${{\pi}\over n}$.

We recall that the contribution to $Z_D'(0)$ from a
(boundary) corner
is
\begin{eqnarray}
A + B\alpha -{1\over 4}\log\alpha +
{1\over{12\alpha}}(\gamma -\log 2 -\log\alpha)
-{1\over{12}}\alpha\log\alpha+J(\alpha) \nonumber
\end{eqnarray}
where the last term has the integral representation
\begin{eqnarray}
\label{Jdefhere}
J(\alpha) =
\int_{0}^{\infty}
{1\over{e^{\mu} -1}}
[{1\over {2\mu}}\coth({x\over{2\alpha}}) -
{{\alpha}\over {4\sinh^2 ({{\mu}\over 2})}}
-{1\over {12}}({1\over{\alpha}}+\alpha)] d\mu
\end{eqnarray}
and have included an undetermined term linear in the
angles ($B\alpha$), and an integration constant, that we
will not need explicitly ($A$).

The analysis proceeds by computing the integral for
$\alpha$ being ${1\over n}$. Let $y$ stand for $e^{\mu}$.
Decomposing various rational functions of $y$:
\begin{eqnarray}
\log (y^n-1) &= &\sum_{v=0}^{n-1} \log (y-\lambda_v)
\qquad\qquad \lambda_v = e^{{2\pi i v}\over n}\nonumber \\
{n\over{y^n-1}} &= &\sum_{v=0}^{n-1} {1\over{\lambda_vy-1}}\nonumber \\
{1\over {y-1}}({n\over{y^n-1}} +{n\over 2}) &=
&{1\over{(y-1)^2}} + {1\over{2(y-1)}} -
\sum_{v=1}^{n-1} {{\lambda_v}\over{1-\lambda_v}}
{1\over{y-\lambda_v}}\nonumber \\
\sum_{v=1}^{n-1} {{\lambda_v}\over{1-\lambda_v}}
&= &{{1-n}\over{2}} \nonumber
\end{eqnarray}
we can absorb the terms proportional to ${1\over n}$ into $B$, and
write the modified integral:
\begin{eqnarray}
\label{tildeJdefn}
\tilde J({1\over n}) =
{1\over n}\int_{\epsilon}^{\infty}
[-{1\over {x}}
\sum_{v=1}^{n-1} {{\lambda_v}\over{1-\lambda_v}}
{1\over{y-\lambda_v}} - {{n^2-1}\over{12(y-1)}}]
 dx
\end{eqnarray}
The cut--off $\epsilon$ is introduced for later convenience.
In fact, we can go backwards and express \ref{tildeJdefn} as
\begin{eqnarray}
\label{tildeJdefc}
\tilde J(\alpha ={1\over n}) =
\int_{0}^{\infty}d\mu
{1\over{e^{\mu}-1}}
[({1\over {2\mu}})(\coth({{\mu}\over{2\alpha}})
- \alpha\coth({{\mu}\over{2}})) -
{1\over{12}}({1\over{\alpha}}-\alpha)]
\end{eqnarray}
Since up to simple terms in $\alpha$ that
we keep, \ref{tildeJdefc} and
\ref{Jdefhere} agree on the integers, and both expressions are
analytic in ${1\over{\alpha}}$, then they must agree overall.
We are therefore allowed to substitute for \ref{Jdefhere}
the more symmetric expression \ref{tildeJdefc}.
Continuing the analysis of \ref{tildeJdefn} we use
\begin{eqnarray}
\int_{\epsilon}^{\infty} {{dx}\over x}
{{\lambda_v}\over{y-\lambda_v}} &=&
\int_{n\epsilon}^{\infty} {{dx}\over x} \sum_{p=1}^n \lambda_{vp}
[{{e^{(1-{p\over n})x}}\over{e^x-1}}] \nonumber \\
\int_{\epsilon}^{\infty} {{dx}\over x}
{{e^{(1-a)x}}\over{e^x-1}} &=& {1\over{\epsilon}}
-({1\over 2} -a)(\gamma + \log\epsilon) + \log\Gamma(a)
-{1\over 2}\log (2\pi) + \hbox{small as $\epsilon$}\nonumber \\
\sum_{v=1}^{n-1} [{{\lambda_{vp}}\over{1-\lambda_v}}]
&=& p - {{1+n}\over 2} \nonumber
\end{eqnarray}
and express the integral $\tilde J$ as
\begin{eqnarray}
\label{tildeJdef2}
\tilde J({1\over n}) =
-{1\over n}\sum_{v=1}^{n-1}\sum_{p}^{n}
{{\lambda_{vp}}\over{1-\lambda_v}}
\int_{n\epsilon}^{\infty}
{{dx}\over x}
{{e^{(1-{p\over n})x}}\over{e^x-1}} + {{n^2-1}\over{12n}}\log\epsilon
\end{eqnarray}
which may finally be simplified to
\begin{eqnarray}
\label{tildeJdef3}
\tilde J({1\over n}) =
{{1-n^2}\over{12n}}(\gamma+\log n) -{1\over{4n}}\log n +
{1\over 4}(1-{1\over n})\log(2\pi) + \sum_{p=1}^{n-1}
({1\over 2} - {p\over n})\log\Gamma({p\over n})
\end{eqnarray}
Absorbing further terms proportional to $1\over n$ into $B$,
we have finally
\begin{eqnarray}
\label{finalexpr}
Z'_{1\over n}(0) = \hat A + \hat B{1\over n} +
{{1-n}\over{12}}\log 2 + ({1\over 4} - {1\over{12n}})\log n +
\sum_{p=1}^{n-1}
({1\over 2} - {p\over n})\log\Gamma({p\over n})
\end{eqnarray}

\subsection*{The integration constant}
We need one exact value to fix an integration constant.
For the equlateral triangle, the regularized determinant
is ({\it see} Appendix~11):
\begin{eqnarray}
Z'_{\hbox{equilateral}}(0)
=  {1\over 2}\log\pi -{1\over 6}\log 2 + {2\over 3}\log 3 + {1\over 2}
\log {{\Gamma({1\over 3})}\over {\Gamma({2\over 3})}}
\end{eqnarray}
If we on the other hand use \ref{finalexpr}, we find
\begin{eqnarray}
Z'_{\hbox{equilateral}}(0) = 3Z_{1\over 3}'(0) =
 -{1\over 2}\log 2 + {2\over 3}\log 3 + {1\over 2}
\log {{\Gamma({1\over 3})}\over {\Gamma({2\over 3})}}
+ 3\hat A + \hat B
\end{eqnarray}
so we have the integration constant
\begin{eqnarray}
3\hat A + \hat B = {1\over 3}\log 2 + {1\over 2}\log\pi
\end{eqnarray}
\subsection*{Simplified general expression}
We choose to include the integration constant in the
stricly local contribution to $Z_D'(0)$, by adding
\begin{eqnarray}
(1-\alpha)[{1\over 6}\log 2 + {1\over 4}\log\pi]
\end{eqnarray}
to a boundary corner (and twice this quantity to an interior
corner).
The final expression with an opening angle ${1\over n}$ then reads
\begin{eqnarray}
\label{finalfixed}
Z'_{1\over n}(0) = (1-{1\over n})({1\over 6}\log 2 + {1\over 4}\log\pi)
+
{{1-n}\over{12}}\log 2 + ({1\over 4} - {1\over{12n}})\log n
\nonumber \\
+ \sum_{p=1}^{n-1}
({1\over 2} - {p\over n})\log\Gamma({p\over n})
\end{eqnarray}
For a general opening angle we go back to
\ref{tildeJdefc}, keep track of the various terms proportional to
${1\over n}$ that had been absorbed into $B$, and arrive at
\begin{eqnarray}
\label{totalintegratedbc}
Z'_{\alpha}(0) = {1\over{12}}({1\over{\alpha}} -\alpha)(\gamma -\log 2)
-{1\over{12}}({1\over{\alpha}} + 3 +\alpha)\log\alpha +\tilde J(\alpha)
\end{eqnarray}
Throughout this appendix, we have simplified on the integral $J$, which
is the same for corners on the boundary and in the interior. The
strictly local contribution to $Z_D'(0)$ from a corner in the
interior is thus
\begin{eqnarray}
\label{totalintegratedic}
Z'_{\alpha}(0) = {1\over{6}}({1\over{\alpha}} -\alpha)(\gamma -\log 2)
-{1\over{6}}({1\over{\alpha}} + \alpha)\log\alpha +2\tilde J(\alpha)
\end{eqnarray}

\section* {A9. Asymptotics of $Z_{\alpha}'(0)$}
\label{s:app9}
\renewcommand{\theequation}{A9.\arabic{equation}}
\setcounter{equation}{0}
In this section we investigate $Z_{\alpha}'(0)$ when $\alpha$ is
large or small or close to one.
\subsection*{Large and small $\alpha$}
It is convenient to divide up the terms in $Z_{\alpha}'(0)$
according to whether they are
symmetric or antisymmetric under the transformation
$\alpha\to{1\over{\alpha}}$.
If we take the integral $\tilde J$, it makes sense
to first introduce a finite symmetric alternative
expression
\begin{eqnarray}
\label{JSdef}
J_S(\alpha) =
\int_{\epsilon}^{\infty}
{{dy}\over y} {1\over{e^{y\sqrt{\alpha}}-1}}
{1\over{e^{y\over{\sqrt{\alpha}}}-1}}
-{1\over{2\epsilon^2}} +
({1\over{\sqrt{\alpha}}}+ \sqrt{\alpha}){1\over{2\epsilon}}
+{1\over {12}}(\alpha+3+{1\over{\alpha}})\log\epsilon
\end{eqnarray}
where the limit as $\epsilon\to 0 $ is understood.
Using the elementary integrals
\begin{eqnarray}
\int_{\epsilon}^{\infty} {{dx}\over{e^x-1}}
&=& -\log\epsilon  + {\cal O}(\epsilon)\nonumber \\
\int_{\epsilon}^{\infty} {{dx}\over{e^x-1}} {1\over x}
&=& {1\over{\epsilon}} + {1\over 2}\log\epsilon +
{{\gamma}\over 2} - {1\over 2}\log 2\pi + {\cal O}(\epsilon)\nonumber \\
\int_{\epsilon}^{\infty} {{dx}\over{e^x-1}} {1\over {x^2}}
&=& {1\over{2\epsilon^2}} - {1\over{2\epsilon}} -
{1\over {12}}\log\epsilon +
+ {1\over{12}} - {{\gamma}\over{12}} - \zeta'(-1) +
{\cal O}(\epsilon) \nonumber \\
\int_{\epsilon}^{\infty} {{dx}\over x} {{e^x}\over {(e^x-1)^2}}
&=& {1\over{2\epsilon^2}}
{1\over {12}}\log\epsilon +
+ {1\over{12}} + {{\gamma}\over{12}} + \zeta'(-1)
+ {\cal O}(\epsilon) \nonumber \\
\end{eqnarray}
we can express the difference between $\tilde J$ and $J_S$
as
\begin{eqnarray}
\label{Jdiff}
\tilde J(\alpha) = J_S(\alpha)
+ {1\over{24}}(\alpha+3+{1\over{\alpha}})\log\alpha
+{{1+\alpha}\over 4}(\gamma-\log 2\pi) -
\alpha({{\gamma}\over{12}}+\zeta'(-1))
\end{eqnarray}
and divide up $Z_{\alpha}'(0)$ in symmetric, antisymmetric,
and linear parts:
\begin{eqnarray}
\label{zprimsep}
Z_{\alpha}'(0) &=& [J_S(\alpha) + {{\gamma}\over{12}}({1\over{\alpha}}
+3+\alpha) - {1\over 4}\log 2\pi] \nonumber \\
&\, & - [{1\over{12}}({1\over{\alpha}}-\alpha) \log 2
+{1\over{24}}({1\over{\alpha}} +3+\alpha)\log\alpha] \nonumber \\
&\, & -\alpha[{1\over 4}\log 2\pi +\zeta'(-1)]
\end{eqnarray}
The behaviour of $J_S$ as $\alpha$ turns to zero is found
expanding the integral in \ref{JSdef}, and subtracting the
terms in the expansion divergent with $\epsilon$:
\begin{eqnarray}
\label{JSexpansion}
J_S(\alpha) &\sim_{\alpha\to 0}&
{1\over{24}}({1\over{\alpha}}
+3+\alpha)\log\alpha
+ ({{1-\gamma}\over{12}}-\zeta'(-1)){1\over{\alpha}} \nonumber \\
&\, & + {1\over 4}(\log 2\pi -\gamma) +
\sum_{n=3}^{\infty} {{\zeta(n)B_{n+1}}\over{n(n+1)}}\alpha^n
\end{eqnarray}
and from this we find
\begin{eqnarray}
\label{zprimexpansion}
Z_{\alpha}'(0) &\sim_{\alpha\to 0}&
{1\over{\alpha}}({1\over{12}}(1-\log 2)-\zeta'(-1))
+ \alpha ({{\gamma+\log 2}\over{12}}-{1\over 4}\log 2\pi-\zeta'(-1))
\nonumber \\
&\, & +\sum_{n=3}^{\infty} {{\zeta(n)B_{n+1}}\over{n(n+1)}}\alpha^n
\end{eqnarray}

Using the symmetry under $\alpha\to{1\over{\alpha}}$ we
have the asymptotic expansion for large $\alpha$:
\begin{eqnarray}
\label{zprimexpansion2}
Z_{\alpha}'(0) &\sim_{\alpha\to \infty}&
-{1\over{12}}({1\over{\alpha}} +3+\alpha)\log\alpha +
\alpha ({1\over{12}}(1-\log 2)-{1\over 4}\log 2\pi-2\zeta'(-1))
\nonumber \\
&\, & + {1\over{\alpha}} ({{\gamma-\log 2}\over{12}})
+\sum_{n=3}^{\infty} {{\zeta(n)B_{n+1}}\over{n(n+1)}}\alpha^{-n}
\end{eqnarray}

\subsection*{Development of $Z_{1+\epsilon}'(0)$ }
It is convenient to go over to the first definition
of the integral $J(\alpha)$ and write
\begin{eqnarray}
Z_{\alpha}'(0) =
{1\over{12}}({1\over{\alpha}} -\alpha)(\gamma -\log 2)
-{1\over{12}}({1\over{\alpha}} + 3 +\alpha)\log\alpha +J(\alpha)
-\alpha \Delta J
\nonumber
\end{eqnarray}
where $J$ has the integral representation
\begin{eqnarray}
J(\alpha) =
\int_{0}^{\infty}
{1\over{e^{\mu} -1}}
[{1\over {2\mu}}\coth({{\mu}\over{2\alpha}}) -
{{\alpha}\over {4\sinh^2 ({{\mu}\over 2})}}
-{1\over {12}}({1\over{\alpha}}+\alpha)] d\mu  \nonumber
\end{eqnarray}
and the difference has the integral representation
\begin{eqnarray}
\Delta J &=&
\int_{0}^{\infty} dx
[ {1\over x}({1\over{(e^x-1)^2}} + {1\over{2(e^x-1)}})
-{{e^x}\over{(e^x-1)^3}} - {1\over{6(e^x-1)}}] \nonumber \\
&=& -{1\over 6}\gamma - {5\over{24}} +{1\over 4}\log(2\pi) + \zeta'(-1)
\end{eqnarray}
The derivative of $J(\alpha)$ can be expanded around $\alpha=1$, and
the successive terms evaluated in Mathematica, which gives
\begin{eqnarray}
J'(1+\epsilon) =
-{1\over 36} \epsilon + {1\over 16} \epsilon^2 + {\cal O}(\epsilon^3)
\end{eqnarray}
and putting the various terms together we have for a corner
on the boundary:
\begin{eqnarray}
\label{zprimexpb}
Z_{1+\epsilon}'(0) &=&
({1\over 6}\log 2 - {5\over{24}} -
{1\over 4}\log(2\pi) - \zeta'(-1))\epsilon \nonumber \\
&\quad & + ({14\over 72} +{{\gamma-\log 2}\over{12}})\epsilon^2
+ (-{29\over 144} - {{\gamma-\log 2}\over{12}})\epsilon^3
+ {\cal O}(\epsilon^4)
\end{eqnarray}
The expansion for a corner in the interior is easily obtained
by adding the expansion of ${1\over 4}\log(1+\epsilon)$, and
doubling that result.

\section* {A10. Special cases}
\label{s:app10}
\renewcommand{\theequation}{A10.\arabic{equation}}
\setcounter{equation}{0}
\subsection*{Triangles}
For triangles one can choose a convenient parametrization
in terms of the Schwarz--Christoffel transformation that maps
a point ($z$) in the triangle to a point ($\omega$) in
the upper complex plane:
\begin{equation}
{{d\omega}\over{dz}}
= \omega^{1-\alpha_0}(1-\omega)^{1-\alpha_1}
\end{equation}
This fixes the area in terms of the angles to be
\begin{equation}
\label{normareaa10}
Area(\alpha_1,\alpha_2,\alpha_{3}) = {{\pi}\over 2}
{{\Gamma(\alpha_1)\Gamma(\alpha_2)\Gamma(\alpha_{3})}\over
{\Gamma(1-\alpha_1)\Gamma(1-\alpha_2)\Gamma(1-\alpha_{3})}}
\end{equation}
One corner (with opening angle $\pi\alpha_0$) will
map to the origin in the $\omega$--plane, one corner
($\pi\alpha_1$) to one, and the last corner maps to infinity.
The parametrization is uniform over the space of triangles,
in the sense that the branchpoints do not move.
Hence the third term in \ref{varycorner} is identically
zero.
The logarithm of the distance between the origin and one is
zero, hence the second terms in \ref{varycorner} are also
zero. All we have left is then the strictly local contribution,
so for this choice of the normalized areas, we have
for triangles:
\begin{equation}
Z_T'(0) = \sum_{p=1,2,3} Z'_{\alpha_p}(0)
\end{equation}
An integration constant was included in the definition of
$Z'_{\alpha}(0)$ in Appendix~8.

By numerically solving for about the first one thousand
eigenvalues of the Laplacian in isosceles
triangles, and then estimating the analytic continuation
of the zeta functions, Luck found that the quotient
\begin{equation}
\zeta_T = {{Z_T'(0)}\over{Z_T(0)}}
\end{equation}
varies surprisingly little over the triangles (still
taking the area \ref{normareaa10})\cite{JMLprive}.
We can express this $\zeta_T$ as
\begin{equation}
\zeta_T = {{\sum_{p=1,2,3} Z'_{\alpha_p}(0)}\over{\sum_{p=1,2,3}
Z_{\alpha_p}(0)}}
\end{equation}
It has a maximum for the equilateral triangle, where
it equals $4.591151\ldots$

The minimum of $\zeta_T$ is obtained as an angle tends to zero,
and the value follows from the asymptotic expansion \ref{zprimexpansion}:
\begin{eqnarray}
\label{zprimelimit}
\hbox{Limit}_{\alpha\to 0} {{Z_{\alpha}'(0)}
\over{Z_{\alpha}(0)}} = 2(1-\log 2)-24\zeta'(-1)
= 4.583813\ldots
\end{eqnarray}

\subsection*{Polygons}
In this case it is convenient to take a parametrization
where the interior of the polygon ($z$) is mapped onto
the interor of a circle ($u$) by a transformation that satisfies
\begin{equation}
{{du}\over{dz}}
= \prod_v (u-e^{i\phi_v})^{1-\alpha_v}
\end{equation}
where $e^{i\phi_v}$ are the branchpoints, and the interior
angles are $\pi\alpha_v$.
The important special case is a {\it regular} $n$--polygon,
$P_n$, that
can be mapped to the unit circle by a transformation that
satisfies
\begin{equation}
{{du}\over{dz}}
= \prod_{v=0}^n (u-e^{{2\pi i v}\over n})^{{2\over n}}
= (u^n-1)^{{2\over n}}
\end{equation}
This parametrization fixes the radius of the circumscribed
circle:
\begin{equation}
\label{normalradius}
R_n = {{\Gamma(1+{1\over n})\Gamma(1-{2\over n})}
	\over{\Gamma(1-{1\over n})}}
\end{equation}
and the area, which is proportional to $R_n^2$.
We can now choose a family of polygons that smoothly
interpolate between a regular $m$--polygon (at $a=0$),
and a regular $mn$--polygon (at $a=1$):
\begin{equation}
{{du}\over{dz}}
= (u^m-1)^{{2\over n}(1-a)}(u^{mn}-1)^{{2\over {mn}}a}
\end{equation}
which is convenient, since the parametrization
is then uniform in
the family. The variation of $Z_D'(0)$ will therefore
be determined by the strictly local terms, and by the second
term in \ref{varycorner}. Let us express $Z_D'(0)$
as a function of the parameter $a$ in the family,
and write out these
two terms in \ref{varycorner} in more detail:
they give the following contribution
to ${d\over{da}}Z_{D(a)}'(0)$ from the corner $v$, which
is mapped to the branchpoint $e^{{2\pi i v}\over{mn}}$;
\begin{eqnarray}
{d\over{da}}Z_{D(a)}'(0)]_{\hbox{from corner $v$}}
= {d\over{da}}Z_{\alpha_v(a)}'(0) +
Z_{\alpha_v(a)}(0) {{d\lambda_v(a)}\over{da}}  \nonumber \\
{{d\lambda_v(a)}\over{da}} = - \sum_{v'\neq v}
{{d\alpha_{v'}(a)}\over{da}}\log |u_{v'}-u_v|^2 \nonumber
\end{eqnarray}
which can be rewritten
\begin{eqnarray}
\label{varynotmanifest}
{d\over{da}}Z_{D(a)}'(0)]_{\hbox{from corner $v$}}
= {d\over{da}}[Z_{\alpha_v(a)}'(0) +
{1\over{12}}({1\over{\alpha_v(a)}} - \alpha_v(a))\lambda_v(a)]
\nonumber \\
+{{\lambda_v(a)}\over 6}{{d\alpha_v(a)}\over{da}}
\end{eqnarray}
We have to differ between whether $n$ does or does not divide
$v$. For the first case
\begin{eqnarray}
\lambda_v(a) &=& -{2\over m}(1-a)\log{{u^m-1}\over{u-1}}|_{u=1}
-{2\over{mn}} a\log{{u^{mn}-1}\over{u-1}}|_{u=1} \nonumber \\
&=&-{2\over m}(1-a + {a\over n})\log m - {{2a}\over{mn}}\log n
\qquad n \mid v
\end{eqnarray}
while for the second
\begin{eqnarray}
\lambda_v(a) = -{2\over m}(1-a)\log |e^{{2\pi i v}\over n} -1| -
{{2a}\over{mn}}\log mn
\qquad n \not\mid v
\end{eqnarray}
The expression \ref{varynotmanifest} contains one part
which is a total variation, and an extra term. Let us first
do the second, summed over the corners $v$:
\begin{eqnarray}
{1\over 6}\sum_v \lambda_v(a) {{d\alpha_v(a)}\over{da}} &=&
{1\over 6} \sum_{n\mid v}
(-{2\over m}(1-a)\log m - {2\over{mn}}\log mn)({2\over m}(1-{1\over n})
\nonumber \\
&+& {1\over 6} \sum_{n\not\mid v}
(-{2\over m}(1-a)\log |e^{{2\pi i v}\over n} -1| - {{2a}\over{mn}}\log mn)
(-{2\over{mn}}) \nonumber
\end{eqnarray}
which can eventually be simplified to
\begin{eqnarray}
-{2\over 3}{{1-a}\over m}\log m +
 {2\over 3}{{1-a}\over {mn}}\log mn \nonumber
\end{eqnarray}
from which we have
\begin{eqnarray}
\int_0^1 da {1\over 6}\sum_v \lambda_v(a) {{d\alpha_v(a)}\over{da}}
= {1\over 3}({1\over{mn}}\log mn - {1\over m}\log m)
\end{eqnarray}
The total variation in \ref{varynotmanifest} is summed over
the corners, and gives in integrated form:
\begin{eqnarray}
nZ_{1 - {2\over n}}'(0) - {{2(n-1)}\over{3n(n-2)}} \log n
+ \hbox{Const.}
\end{eqnarray}
We are now ready to write down $Z_{P_n}'(0)$ for
a regular polygon with $n$ corners and radius of circumscribed
circle $R$:
\begin{eqnarray}
Z_{P_n}'(0) &=& Z_{P_n}(0)\log{{R^2}\over{R_n^2}} -
{1\over{3(n-2)}}\log n + n Z_{1 - {2\over n}}'(0)
\end{eqnarray}
where $Z_{P_n}(0) = {{n-1}\over{6(n-2)}}$ and the normal
radius is as in \ref{normalradius}. The integration constant
turns out to be zero, since it has already been
incorporated in the definition of $Z_{\alpha}'(0)$.
\subsection*{Square and disc}
For the square we obtain
\begin{eqnarray}
Z_{P_4}'(0) &=&
{1\over 2}\log({{\Gamma({3\over 4})4R}\over{\sqrt{\pi}\Gamma({1\over 4})}})
- {1\over 6}\log 4 + 4({1\over 8}\log\pi + {5\over{24}}\log 2) \nonumber \\
&=& {1\over 2}\log{{\Gamma({3\over 4})}\over{\Gamma({1\over 4})}}
+ {1\over 4}\log\hbox{Area} + {1\over 4}\log\pi + {5\over 4}\log 2
\end{eqnarray}
which agrees with the exact result from \ref{squareexact}.

The disc is obtained as the limit when $n$ tends to infinity.
We then find:
\begin{eqnarray}
Z_{P_{\infty}}'(0) &=&
{1\over 3}\log R - 2{d\over{d\alpha}}Z_{\alpha}'(0)|_{\alpha=1} \nonumber \\
&=& {1\over 3}\log R + {5\over 12} + {1\over 2}\log\pi + {1\over 6}\log 2
+ 2\zeta'(-1)
\end{eqnarray}
which agrees with \ref{discexact}.

\section* {A11. Integrable domains.}
\label{s:app11}
\renewcommand{\theequation}{A11.\arabic{equation}}
\setcounter{equation}{0}
In this appendix we collect the cases known to us,
where one can directly deduce the derivative of the
zeta function at zero.

In a rectangle with side lengths $A$ and $B$, the eigenvalues
of the Laplacian with Dirichlet boundary conditions are
\begin{equation}
\label{eigenvaluesrectangle}
\lambda_{mn} = \pi^2({{m^2}\over{A^2}} + {{m^2}\over{A^2}})
\end{equation}
The zeta function around the origin is
\begin{eqnarray}
Z_{\hbox{rectangle}}(0) = &{1\over 4} \nonumber \\
Z_{\hbox{rectangle}}'(0) =
& {1\over 4}\log (AB) - \log [2^{-{1\over 2}} ({B\over A})^{1\over 4}
\eta(q)] \nonumber
\end{eqnarray}
where $\eta$ is the modular form of Dedekind:
\begin{eqnarray}
\eta(q) = q^{1\over{24}}\prod_{m=1}^{\infty} (1-q^m)
\qquad q = e^{-2\pi\sqrt{{B\over A}}}
\end{eqnarray}
For the square, we have the simpler expression
\begin{eqnarray}
\label{squareexact}
Z_{\hbox{square}}'(0) =
{1\over 4}\log A^2 + {1\over 4}\log
[\pi 2^5{{\Gamma^2({3\over 4})}\over{\Gamma^2({1\over 4})}}]
\end{eqnarray}

Three triangles tile the plane by reflections in the side:
the equilateral (${1\over 3},{1\over 3},{1\over 3}$);
the bisected equilateral
${1\over 2},{1\over 3},{1\over 6}$, and
the right angle isosceles (${1\over 2},{1\over 4},{1\over 4}$).
In these domains one can solve for the eigenmodes of the
Laplacian by superposition of plane waves \cite{Itz,IL},
We normalize the areas in terms of the side lengths ($a$) of the
equilateral, the sidelengths of the legs in the right angle
isosceles,
and the sidelength of the longest side in the
bisected equilateral:
\begin{eqnarray}
\label{eigenvaluestriangles}
E_{({1\over 3},{1\over 3},{1\over 3}),n,m} &=&
({{4\pi}\over{3a}})^2(n^2+m^2-nm)\qquad  n>m>0\nonumber \\
E_{({1\over 2},{1\over 4},{1\over 4}),n,m} &=&
({{\pi}\over{a}})^2(n^2+m^2)\qquad  n>m>0\nonumber \\
E_{({1\over 2},{1\over 3},{1\over 6}),n,m} &=&
({{4\pi}\over{3a}})^2(n^2+m^2+nm)\qquad  n>m>0 \nonumber
\end{eqnarray}
The corresponding zeta functions can be written in terms
of the Riemann zeta-function and
Dirichlet $L$-series\cite{IL},
which may in turn be resolved into sums of zeta
functions of Hurwitz $H(x,s)$ with different arguments $x$.
Using
\begin{eqnarray}
\label{Hurwitzzeta}
H(x,s) = \sum_{k=0}^{\infty} {1\over{(x+k)^s}} \nonumber \\
H(x,0) = {1\over 2} -x \qquad\qquad {d\over{ds}}H(x,s)|_{s=0}
= \log {{\Gamma(x)}\over{\sqrt{2\pi}}} \nonumber
\end{eqnarray}
and the normal areas
determined
by the representation of the Schwarz--Christoffel transformation;
\begin{eqnarray}
\label{intareas}
A(\alpha_0,\alpha_1,\alpha_{\infty}) &= {{\pi}\over{2}}
{{\Gamma(\alpha_0)\Gamma(\alpha_1)\Gamma(\alpha_{\infty})}
\over {\Gamma(1-\alpha_0)\Gamma(1-\alpha_1)\Gamma(1-\alpha_{\infty})}}
\end{eqnarray}
one finds after some
algebra the following expressions for the determinants:
\begin{eqnarray}
\label{intzetas1}
Z_{({1\over 3},{1\over 3},{1\over 3})}'(0) &=&
{1\over 3}\log {{\hbox{Area}}\over{A_{({1\over 3}, {1\over 3},
{1\over 3})}}} +
\log {{\Gamma^{1\over 2}({1\over 3})\pi^{1\over 2}
       3^{2\over 3} 2^{-{1\over 6}}}\over{\Gamma^{1\over 2}({2\over 3})}} \\
Z_{({1\over 2},{1\over 4},{1\over 4})}'(0) &=&
{3\over 8}\log {{\hbox{Area}}\over{A_{({1\over 2}, {1\over 4},
{1\over 4})}}} +
\log {{\Gamma^{1\over 2}({1\over 4})\pi^{1\over 2}2^{7\over 8}}
\over{\Gamma^{1\over 2}({3\over 4})}} \\
Z_{({1\over 2},{1\over 3},{1\over 6})}'(0) &=&
{5\over 12}\log {{\hbox{Area}}\over{A_{({1\over 2}, {1\over 3},
{1\over 6})}}} +
\log {{\Gamma({1\over 3})3^{11\over 24} 2^{2\over 9}\pi^{1\over 2}}
\over{\Gamma({2\over 3})}}
\end{eqnarray}

The eigenvalues of the Laplacian in on the upper half sphere
of radius one
are $l(l+1)$. If one imposes Dirichlet
boundary conditions at the equator, one selects out the
eigenmodes odd under reflection in a plane through the
equator. Each eigenvalue will then be $l$ times degenerate.
The zeta function of this spectrum is
\begin{equation}
Z_{\hbox{hemisphere}}(s) = \sum_{l=1}^{\infty} {l\over{(l(l+1))^s}}
\nonumber
\end{equation}
from which one can derive \cite{Weisberger}
\begin{equation}
\label{hemispehereexact}
Z_{\hbox{hemisphere}}'(0) = 2\zeta'(-1) + {1\over 2}\log 2\pi - 4
\end{equation}
The hemisphere can be mapped conformally to a disc with radius
$R$. The difference of the regularized determinants on
the hemisphere and on the disc can then be evaluated by
computing the Liouville action (\ref{Liouvilleaction})
of the conformal factor\cite{Weisberger}:
\begin{equation}
\label{discexact}
Z_{\hbox{disc}}'(0) = {1\over 6}\log 2 + {1\over 2}\log\pi +
{1\over 3}\log R + 2\zeta'(-1) + {5\over{12}}
\end{equation}
Equation \ref{discexact} has also been checked numerically
to great accuracy by computing the eigenvalues from the
zeros of Bessel functions, and directly investigating the
analytical continuation of the zeta function\cite{JMLprive}.

\end{document}